%% file: main.tex
\documentclass[12pt]{article}

\usepackage[
  left=0.8in,
  right=0.8in,
  top=0.6in,
  bottom=0.8in
]{geometry}

\usepackage[utf8]{inputenc}
\usepackage[T1]{fontenc}
\usepackage[english]{babel}
\usepackage{nameref}
\usepackage{todonotes}
\usepackage{graphicx}
\usepackage{subcaption}
\usepackage[font=small,labelfont=bf]{caption}
\usepackage{ragged2e}
\usepackage{float}
\usepackage{adjustbox}

\usepackage{amsmath}
\usepackage{amssymb}
\usepackage{mathtools}
\usepackage[version=4]{mhchem}
\usepackage{physics}
\usepackage{bigints}
\usepackage{setspace}
\usepackage{booktabs}
\usepackage{tabularx}
\usepackage{makecell}

\usepackage[super,sort&compress,numbers]{natbib}
\usepackage[colorlinks=true,allcolors=black]{hyperref}

\usepackage{xcolor}
\usepackage[normalem]{ulem}
\usepackage{changes}
\usepackage{todonotes}

\definecolor{addedcolor}{rgb}{0,0.5,0}
\definecolor{deletedcolor}{rgb}{1,0,0}
\setdeletedmarkup{\textcolor{deletedcolor}{\sout{#1}}}
\setaddedmarkup{\textcolor{addedcolor}{#1}}

\usepackage{newunicodechar}
\newunicodechar{​}{} 
\DeclareUnicodeCharacter{2009}{\,}
\DeclareUnicodeCharacter{2212}{\ensuremath{-}}

\title{\textbf{Amorphous silicon metasurfaces encode light-driven catalytic activity}}
\author{%
Elif Nur Dayi$^{1}$,
Priscila Vensaus$^{1}$,
Milad Sabzehparvar$^{1}$,
Michele Puppin$^{2}$,
Omer Can Karaman$^{1}$,
Giulia Tagliabue$^{1,*}$\\[0.75em]
{\small
\textsuperscript{1}Laboratory of Nanoscience for Energy Technologies (LNET), STI,\\
École Polytechnique Fédérale de Lausanne, 1015 Lausanne, Switzerland\\[0.75em]
\textsuperscript{2}Lausanne Centre for Ultrafast Science (LACUS), STI,\\
École Polytechnique Fédérale de Lausanne, 1015 Lausanne, Switzerland\\[0.75em]
\textsuperscript{*}Correspondence: giulia.tagliabue@epfl.ch
}%
}

\begin{document}
\setstretch{1.15}
\begin{titlepage}
\centering

\vspace*{1cm}

{\Large\bfseries Metaphotonic Catalysis: Amorphous silicon \\metasurfaces encode 
photochemical activity\par}

\vspace{1cm}

{\normalsize
Elif Nur Dayi$^{1}$, Priscila Vensaus$^{1}$, Milad Sabzehparvar$^{1}$,\\
Michele Puppin$^{2}$, Omer Can Karaman$^{1}$, Giulia Tagliabue$^{1,*}$
\par}

\vspace{0.6cm}

{\small
$^{1}$Laboratory of Nanoscience for Energy Technologies (LNET), STI,\\
École Polytechnique Fédérale de Lausanne, 1015 Lausanne, Switzerland
\par}
\small $^{2}$Lausanne Center for Ultrafast Science (LACUS), STI,\\
\small École Polytechnique Fédérale de Lausanne, 1015 Lausanne, Switzerland\\[0.75em]
\vspace{0.2cm}

{\small
$^{*}$Correspondence: giulia.tagliabue@epfl.ch
\par}

\vspace{1cm}


\vspace{1cm}

\begin{abstract}
\small

Solar-to-fuel conversion can benefit from photoelectrodes with engineered light-matter interactions, yet most nanostructured designs provide limited control over the spatial and spectral distribution of photochemical activity. Here, we present an all-dielectric amorphous-silicon metasurface photoelectrode that confines resonant light-matter interactions within a 220-nm-thick active layer. Tunable Mie-type and guided-mode resonances spectrally encode chemical reactivity and produce absorptance above 80\% near the silicon band-edge, compared with less than 30\% for an unpatterned film. The metasurface functions as the light absorber, carrier-transport layer, and catalytic interface without an added co-catalyst or engineered passivation layer.

Operando photo-scanning electrochemical microscopy reveals wavelength- and structure-dependent redox activity and a tenfold enhancement in internal quantum efficiency near the silicon band-edge relative to planar films. Power-dependent measurements support a photon-driven rather than nonlinear photothermal origin, while surface-sensitive ultrafast transient-reflectivity measurements probe the underlying carrier dynamics. Light-coupled scanning electrochemical cell microscopy shows hydrogen-evolution enhancements of up to 21-fold under photocatalytic conditions and 15-fold under photoelectrochemical bias, corresponding to 11.2-fold and 7.7-fold enhancements after accounting for the estimated surface-area increase. The metasurfaces remain stable during more than 10 hours of immersion and prolonged laser illumination, establishing amorphous silicon as a stable platform for resonantly programmed photocatalysis and solar-fuel generation.

\end{abstract}

\end{titlepage}

\input{sections/sections.tex}


\bibliographystyle{unsrt}
\bibliography{ref}
\newpage
 \input{sections/data.tex}
\input{sections/supportinginfo.tex}
\input{sections/acknowledgements.tex}
\input{sections/contributions.tex}
\input{sections/comp.tex}


\newpage
\input{sections/toc}

\end{document}

%% file: sections/sections.tex
\section{Main} \label{sec:Intro}

Photocatalytic (PC) and photoelectrochemical (PEC) systems offer promising routes for converting solar energy into chemical fuels\cite{segev_2022_2022}. A central challenge in solar fuel generation is the simultaneous optimization of light absorption, charge-carrier transport, and interfacial catalytic activity within a simple device architecture \cite{peng_progress_2025}. Planar semiconductor photoelectrodes remain fundamentally constrained by the trade-off between efficient optical absorption and carrier extraction. Typically, thick absorbers are required to efficiently harvest sunlight, yet increased charge transport distances exacerbate recombination losses and limit interfacial conversion efficiencies.
Nanostructured semiconductor electrodes have been widely explored to address these limitations by increasing surface area, enhancing light trapping, and reducing carrier transport lengths \cite{andrei_nanowire_2023,roh_photoelectrochemical_2022,boettcher_photoelectrochemical_2011,choi_sn-coupled_2014}. However, most non-resonant architectures increase absorption primarily through geometric scattering and increased interface density, offering limited control over the spatial localization and spectral selectivity of light-driven processes. As a result, the relationship between optical excitation and local chemical reactivity cannot be deterministically engineered.

The ability to tailor the spectral response near the semiconductor optical band-edge is particularly important for solar-fuel generation since absorption in this region lies at the center of a thermodynamic trade-off in semiconductor light absorbers: photons substantially below the optical gap are absorbed only weakly and contribute little to carrier generation, whereas photons far above the optical gap are absorbed efficiently but incur greater carrier-thermalization losses. Absorption close to the band-edge therefore enables the use of photons that remain sufficiently energetic for charge generation while minimizing excess-energy losses prior to interfacial chemistry. However, unpatterned ultrathin a-Si photoelectrodes typically respond weakly in this spectral region because the absorption coefficient decreases toward the optical band-edge and the optical path length is limited \cite{polman_photonic_2012,chen_progress_2017}.

Dielectric metasurfaces provide an attractive framework for addressing these limitations by enabling resonant control over light–matter interactions at the nanoscale. Composed of subwavelength resonators supporting a rich variety of optical modes, including Mie-type resonances, these structures confine electromagnetic energy within the nanoantenna volume. Such resonant localization can produce near-unity absorption in ultrathin active layers, thereby decoupling optical absorption from carrier transport length scales \cite{hale_perfect_2020,song_laterally_2014}.

These capabilities have motivated the development of all-dielectric metasurfaces as platforms for tailoring photocatalytic responses across the visible and near-infrared spectral ranges \cite{yuan_quasi-bound_2024,mascaretti_designing_2023,cortes_optical_2022}. Huttenhofer \textit{et al.} investigated GaP metasurfaces coated with an ultrathin Pt co-catalyst for the hydrogen evolution reaction \cite{huttenhofer_metasurface_2021}. TiO$_2$ metasurfaces have been explored for photocatalytic Ag reduction, with prior studies employing defect engineering or ion implantation to modify the optical constants, as well as resonant anapole and bound-state-in-the-continuum modes to enhance absorptance \cite{wu_tio2_2019,hu_catalytic_2022,huttenhofer_anapole_2020}. Paudel \textit{et al.} demonstrated resonance-enhanced carrier generation in Au/GaAs hybrid metasurface photoanodes \cite{paudel_metasurface-enhanced_2024}. Collectively, these studies show that resonant nanophotonic effects can enhance optical absorption, photocatalytic activity, and incident photon-to-current conversion efficiencies.

However, metasurface-based photoelectrodes and photocatalysts are still predominantly evaluated using measurements averaged over macroscopic device areas. Moreover, the use of metallic co-catalysts or hybrid plasmonic architectures introduces additional interfaces that can complicate isolation of the intrinsic contribution of the resonant semiconductor surface. Consequently, how wavelength-selective optical resonances translate into spatially resolved chemical activity remains insufficiently understood. All-dielectric metasurfaces composed entirely of a photoactive semiconductor provide a suitable platform for addressing this question by integrating optical resonances and chemical activity within the same material.

Amorphous silicon (a-Si) represents an overlooked yet compelling platform for resonant photocatalysis and photoelectrochemistry. Beyond its low cost, the abundance of silicon, and compatibility with mature semiconductor manufacturing, a-Si exhibits a high refractive index and a tunable optical bandgap in the range of 1.6--1.8 eV \cite{kilner_j_a_photovoltaic_2012,cody_disorder_1981}, enabling strong optical confinement across the visible and near-infrared spectral range. Accordingly, a-Si metasurfaces have become a cornerstone for applications such as nonlinear optics \cite{wang_resonantly_2024,karaman_decoupling_2025,padhy_temperature_2025}, quantum photonics \cite{zhou_metasurface-assisted_2025,karaman_photo-thermally_2026}, and biotechnology \cite{yang_optofluidic_2023,liu_geometry-programmable_2026,pahlevaninezhad_nano-optic_2018}.
Despite these advantages, the role of a-Si in photocatalysis and photoelectrochemistry remains limited \cite{peng_progress_2025,ma_fundamental_2021}, either employing it as a passivating layer \cite{liu_bifacial_2020}  or a light absorber several hundred nanometers thick\cite{li_photoelectrochemical_2019,zhou_breaking_2025}. Along these lines, silicon-based PC and PEC systems have typically relied on planar or otherwise non-resonant architectures, often combined with passivation layers and metallic co-catalysts to stabilize the semiconductor interface and enhance catalytic activity \cite{lin_amorphous_2013,lin_photochemical_2023, lee_scalable_2021,chen_atomic_2011,roh_photoelectrochemical_2022, liu_geometry-programmable_2026, choi_sn-coupled_2014, feng_hydrogen_2018}.  Consequently, resonant nanophotonic control of interfacial photochemistry in silicon-based systems is yet to be systematically studied.

Here, we introduce all-dielectric resonant amorphous silicon metasurfaces as chemically compatible photoelectrodes for photocatalysis and photoelectrochemistry, enabling stable operation and on-chip control of catalytic activity. The metasurfaces support hybridized Mie-type electric and magnetic modes along with guided-mode resonances (GMRs), confining electromagnetic energy within the resonator and underlying film, enabling strong absorptance (> 80\%) near the optical band-edge of amorphous silicon where the intrinsic absorption is weak (<30\%). 

Using spatially resolved scanning photoelectrochemical microscopy, together with optical and structural characterization, we correlate resonant electromagnetic modes with local photoinduced chemical activity. Fast-kinetics redox-couple measurements show that the metasurface geometry can be designed to produce a targeted spectral profile of interfacial photoelectrochemical activity, with up to a tenfold enhancement in internal quantum efficiency near the silicon band-edge relative to planar films. Femtosecond transient-reflection measurements are used to further investigate the carrier dynamics underlying this response. We then employ light-coupled scanning electrochemical cell microscopy to extend this approach to local measurements of the hydrogen evolution reaction under photocatalytic and photoelectrochemical conditions, revealing enhancements of up to 21-fold and 15-fold, respectively. After accounting for the estimated increase in surface area, these values correspond to enhancements of 11.2-fold and 7.7-fold, demonstrating that the improvement extends well beyond geometric surface-area effects. Together, these findings highlight the substantial potential of resonant mode engineering to improve photon utilization and chemical activity in ultrathin semiconductor photoelectrodes.

\section{Amorphous Silicon Metasurfaces} 
\label{sec:aSi metasurface}

\begin{figure}[H]
\centering
\includegraphics[width=\textwidth]{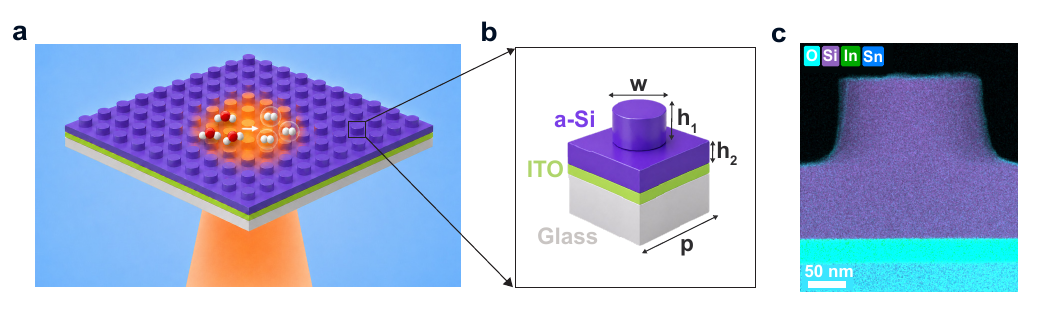}
\captionsetup{font=small}
\caption{\justifying{ Concept of the a-Si metasurface photoelectrode for resonance-driven chemistry.
\textbf{a} Schematic of the a-Si metasurface photoelectrode designed to achieve photonic resonance-driven chemical activity.
\textbf{b} Schematic of the a-Si meta-atom with partially-etched cylindrical nanoresonators on ITO-coated glass. The geometric parameters are diameter ($w$), resonator height ($h_1$), underlying a-Si film thickness ($h_2$), and period ($p$). \textbf{c} Cross-sectional STEM-EDS map of an a-Si nanopillar with $w = 190~\mathrm{nm}$ following the photo-SECM redox measurements. Scale bar: 50 nm.
}}

\label{fig:device}
\end{figure}

We designed an all-dielectric metasurface (MS) photoelectrode that combines strong optical absorption, charge transfer, and catalytic activity in an ultrathin semiconductor without an added co-catalyst or passivation layer (Figure~\ref{fig:device}a). The device comprises a square array of partially etched p-type amorphous silicon (a-Si) nanoresonators on indium tin oxide (ITO)-coated glass. The metasurfaces were fabricated from nominally 220-nm-thick sputtered a-Si films patterned into square lattices with a fixed period of $p = 325~\mathrm{nm}$ and resonator diameters ranging from $w = 130$ to $230~\mathrm{nm}$, as shown in Figure~\ref{fig:device}b. The optical response was tuned by varying the resonator diameter ($w$), etch depth ($h_1$), and residual film thickness ($h_2$).

The buried ITO serves as both a transparent hole-collecting layer and an electrical back contact during photoelectrochemical measurements while remaining fully encapsulated beneath the a-Si. Consequently, the measured photocurrent originates exclusively from photogenerated carriers within the resonant silicon structures. Cross-sectional TEM and EDS mapping confirm the multilayer architecture and elemental composition of the fabricated metasurface (Figure~\ref{fig:device}c).

We next evaluated how the fabricated meta-atom geometry shapes the optical response of the a-Si photoelectrode. To optimize the metasurface and identify the optical modes responsible for the absorption enhancement, we combined optical spectroscopy with full-wave electromagnetic simulations. Experimentally determined optical constants of the sputtered a-Si and ITO layers, obtained by spectroscopic ellipsometry, were used as inputs to the model (Supporting Figure 3a,b), with fabrication, measurement, characterization, and modelling details provided in \nameref{sec:Methods}.

\begin{figure}[H]
\centering
\includegraphics[width=1\textwidth]{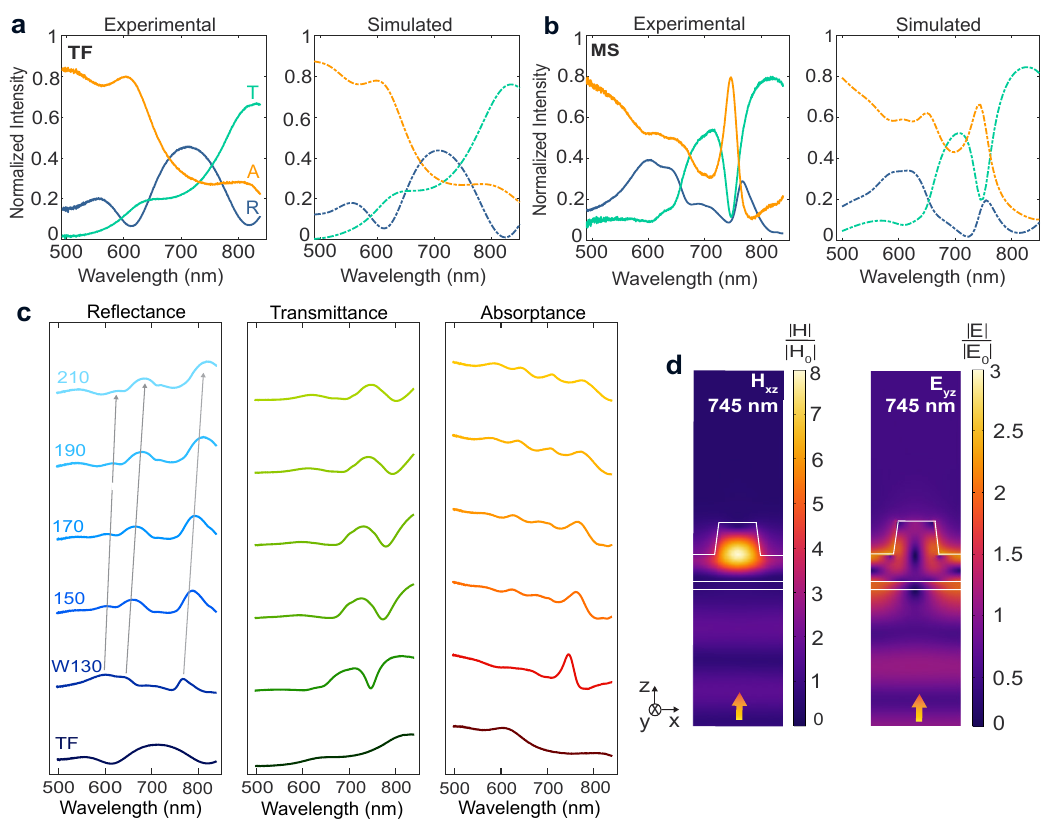}
\captionsetup{font=small}
\caption{\justifying{ \textbf{Optical characterization of the thin film and metasurface photoelectrodes.}
\textbf{a} Experimental and simulated normalized reflectance ($R$), transmittance ($T$), and absorptance ($A$) spectra for the unpatterned thin film (TF) and \textbf{b} for a metasurface (MS) with $w = 130~\mathrm{nm}$, $h_1 = 120~\mathrm{nm}$, $h_2 = 100~\mathrm{nm}$, and $p = 325~\mathrm{nm}$. Solid and dashed curves indicate experiment and simulation, respectively.
\textbf{c} Experimental spectra for the TF and metasurfaces with varying resonator diameters; gray arrows guide the eye for the geometry-dependent shifts of the Mie- and GMR-dominated resonances.
\textbf{d} Simulated $H_xz$ and $E_yz$ field profiles at the near-infrared resonance wavelength of 745 nm, showing guided-mode coupling and field enhancement near the nanopillar-underlying film interface.
}}

\label{fig:absorption}
\end{figure}

Figure \ref{fig:absorption}a compares experimental and simulated normalized reflectance ($R$), transmittance ($T$), and absorptance ($A$) spectra for the unpatterned a-Si thin film (TF) under bottom illumination in air. Bottom illumination was used to match the geometry of the photoelectrochemical measurements. Absorptance was calculated as $A = 1 - R - T$. Experiment and simulation agree strongly across the measured wavelength range, including the Fabry--Perot interference fringes of the unpatterned thin film, whose positions are set by the multilayer thicknesses.

We next focused on a metasurface with $w = 130~\mathrm{nm}$, $h_1 = 120~\mathrm{nm}$, $h_2 = 100~\mathrm{nm}$, and $p = 325~\mathrm{nm}$, which was selected for its strong near-band-edge absorption feature (Figure \ref{fig:absorption}b). The simulations include a 40-nm sidewall taper to account for fabrication imperfections observed by cross-sectional STEM-EDS.  

The optical bandgap (Tauc gap) of the sputtered a-Si, commonly associated with the mobility gap in amorphous semiconductors, was estimated by Tauc analysis \cite{cody_disorder_1981,tauc_optical_1966}. A linear fit yielded $E_g = 1.79~\mathrm{eV}$ with $R^2 = 0.999$ (Supporting Figure~3c), corresponding to a wavelength of approximately 692~nm, consistent with previously reported values \cite{kilner_j_a_photovoltaic_2012,knief_disorder_1999}. The simulations reproduce the experimental optical response, including absorptance peaks near 535, 600, and 745~nm. Minor discrepancies, including small spectral shifts and reflectance deviations below 10\%, likely arise from fabrication-induced structural disorder, variations in the optical constants of the sputtered a-Si, and reduced detector sensitivity above 800~nm. The overall agreement enables reliable identification of the resonant modes responsible for the enhanced absorption.

The metasurface optical response arises from several contributions: Mie-type resonances of the a-Si nanopillars, guided-mode resonances enabled by the partially etched architecture, and Fabry--Perot interference from the multilayer stack. The Mie-type modes arise from the localized resonant response of the nanopillars and therefore depend primarily on the resonator geometry, particularly $w$ and $h_1$. By contrast, the guided-mode resonances arise from coupling between the periodic nanopillar array and the underlying a-Si film. The nanopillar lattice provides the in-plane momentum required to couple normally incident light into the guided mode, while the underlying film provides the leaky waveguiding pathway\cite{wang_theory_1993}. These modes of different nature hybridize, and tuning the geometric parameters such as $w$, $p$ and $h_2$ allows changing the coupling strength between the modes and spectral tuning. To compare multiple geometries under identical fabrication conditions and follow the evolution of the resonances with meta-atom geometry, we patterned arrays with different resonator diameters on the same device. Figure~\ref{fig:absorption}c shows stacked experimental reflectance, transmittance, and absorptance spectra for the unpatterned thin film and metasurfaces with $w = 130$--$210~\mathrm{nm}$, with the overlaid spectra provided in Supporting Figure 4. The resonant features redshift with $w$, consistent with geometry-dependent tuning of the metasurface modes. Having introduced the resonant optical landscape of the metasurface photoelectrode, we next examine how these modes translate into photoelectrochemical response.

\section{Photo-SECM measurements for redox reactions} 
\label{sec:secm_redox}

\begin{figure}[h!]
\centering
\includegraphics[width=17cm]{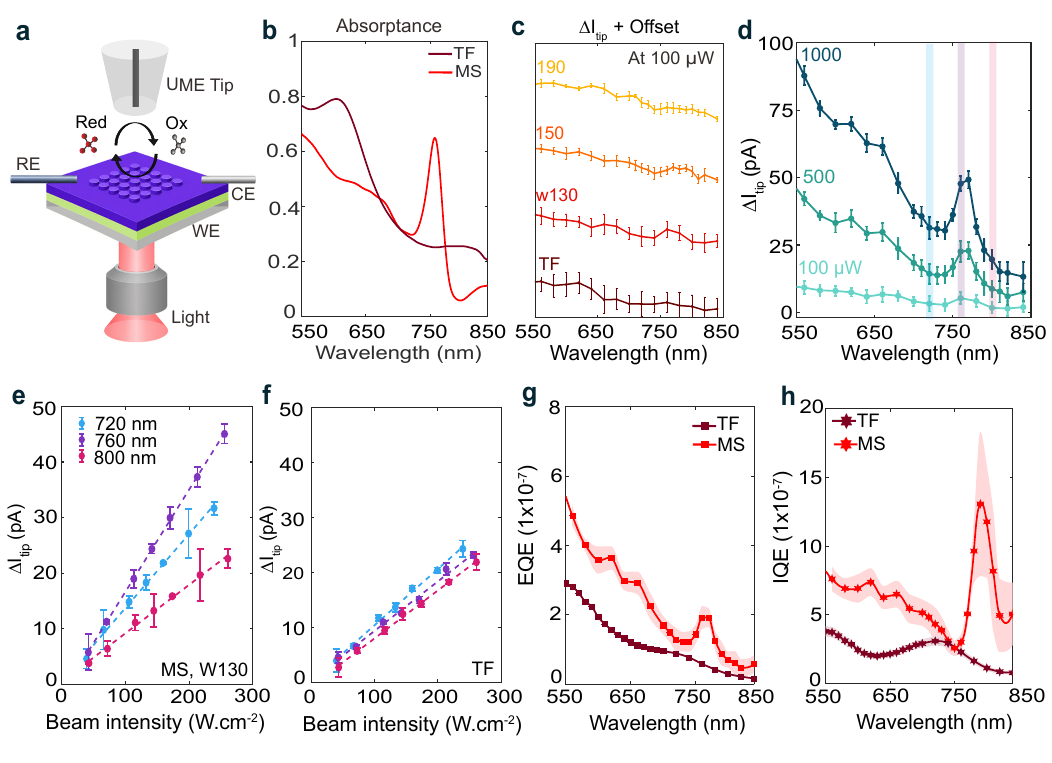}
\caption{\justifying{
\textbf{Photo-SECM measurements of resonant redox activity in a-Si metasurfaces.}
\textbf{a} Bottom-illumination photo-SECM configuration for ferro/ferricyanide redox cycling. In tip-generation/substrate-collection mode, the Pt ultramicroelectrode oxidizes ferrocyanide to ferricyanide, while the illuminated a-Si thin film or metasurface reduces ferricyanide in an aqueous electrolyte containing 0.2~M Na$_2$SO$_4$ and 2~mM ferrocyanide.
\textbf{b} Absorptance spectra of the thin film (TF) and the $w = 130~\mathrm{nm}$ metasurface (MS) in water, showing the guided-mode-resonance dominated feature near 760~nm.
\textbf{c} Wavelength-dependent tip photocurrent $\Delta I_{\mathrm{tip}}$ for the TF and metasurfaces with different resonator widths under 100~$\mu$W illumination. The spectra are vertically offset for clarity and exhibit geometry-dependent features consistent with the corresponding optical resonances. Values are averaged over three illumination cycles, and error bars indicate the standard deviation.
\textbf{d} Power- and wavelength-dependent $\Delta I_{\mathrm{tip}}$ for the $w = 130~\mathrm{nm}$ metasurface, showing an enhanced response near the 760-nm resonance. Vertical shaded bands indicate the wavelengths selected for power-dependent measurements: 720, 760 and 800~nm.
\textbf{e,f} Dependence of $\Delta I_{\mathrm{tip}}$ on beam intensity at 720, 760 and 800~nm for the $w = 130~\mathrm{nm}$ metasurface and TF, respectively. Dashed lines show linear fits, with $R^2 > 0.995$ at all wavelengths, supporting a photon-flux-dependent photoelectrochemical response.
\textbf{g,h} External and internal quantum efficiencies of the metasurface and TF. The metasurface exhibits enhanced EQE and up to approximately tenfold higher IQE near the a-Si band-edge. Each photo-SECM measurement samples an approximately 10-nm illumination bandwidth; the absorptance used to calculate IQE was therefore averaged over the corresponding $\pm 5$-nm interval. Shaded regions show the propagated uncertainty from the photocurrent and absorptance measurements. For the metasurface, the absorptance uncertainty accounts for structural variations across the measured area and an additional 10\% relative uncertainty.
}}
\label{fig:secm_redox}
\end{figure}

Scanning photoelectrochemical microscopy (photo-SECM) enables localized, in situ monitoring of interfacial redox chemistry and quantitative mapping of photochemical activity \cite{mcbrayer_scanning_2024,kiani_distinguishing_2024}. The photo-SECM configuration is schematically shown in Figure \ref{fig:secm_redox}a. Measurements were performed in tip-generation/substrate-collection mode, in which the ultramicroelectrode (UME) tip oxidizes ferrocyanide to ferricyanide, while the illuminated a-Si thin film or metasurface electrode reduces ferricyanide back to ferrocyanide. The resulting faradaic tip current provides a local measure of the photoelectrochemical activity of the illuminated substrate. Figure \ref{fig:secm_redox}b shows the experimentally measured absorptance in water for the thin film and  metasurface geometries introduced in Figure \ref{fig:absorption}a-b. Compared with the absorptance spectra measured in air, the metasurface resonances are redshifted, with the GMR-dominated resonance shifting from 745 to 760 nm, consistent with the simulations shown in Supporting Figure 5.

Figure~\ref{fig:secm_redox}c shows the wavelength-dependent tip photocurrent, $\Delta I_{\mathrm{tip}}$, extracted from chopped-light chronoamperometry at a constant incident power of $100~\mu\mathrm{W}$ for the metasurface geometries shown in Figure ~\ref{fig:absorption}c and Figure ~\ref{fig:secm_redox}b. 
The photo-SECM photocurrent spectra mirror the geometry-dependent resonant behavior identified optically, as further illustrated by the overlaid optical and photo-SECM spectra in Supporting Figure 7. The strongest near-infrared GMR-associated response is observed for (\(w = 130~\mathrm{nm}\)), showing that the metasurface can effectively modulate the interfacial chemical response. 

The metasurface photoelectrode remained stable during prolonged aqueous exposure. After 10 hours of immersion, the $w=130~\mathrm{nm}$ metasurface reproduced the same wavelength-dependent photocurrent response, confirming that the observed spectral pattern was intrinsic to the metasurface rather than a tip-induced artifact (Supporting Figure~6). The optical absorptance spectrum also retained its characteristic resonant features after immersion. Post-reaction cross-sectional STEM-EDS characterization further showed that the nanopillar morphology and multilayer structure remained intact, with no detectable impurity deposits on the surface (Figure~\ref{fig:device}c).

Together, these results show that the photoelectrochemical response of the metasurface can be directed by nanophotonic resonance engineering. In particular, the GMR-dominated near-infrared mode enhances photocurrent generation near the a-Si band-edge, where the unpatterned thin film exhibits weak response, thereby extending photoelectrochemical activity into a spectral region that is otherwise poorly utilized.

We next examined the power dependence of the spectral photocurrent response for the $w = 130~\mathrm{nm}$ metasurface. Figure~\ref{fig:secm_redox}d shows wavelength-dependent $\Delta I_{\mathrm{tip}}$ spectra measured as the incident power was increased from $100$ to $1000~\mu\mathrm{W}$. The resonant features remain fixed in wavelength while becoming progressively more pronounced with increasing power. At the GMR-dominated resonance near 760 nm, $\Delta I_{\mathrm{tip}}$ increases by approximately an order of magnitude across this power range.

To assess whether the enhanced response originates from photon-driven charge generation rather than heating-related effects, we performed wavelength-resolved power-dependent measurements at three positions across the near-infrared resonance: 720 nm on the high-energy shoulder, 760 nm at the resonance maximum, and 800 nm on the low-energy off-resonance tail. Wavelengths further into the near-infrared were not used, as the available laser power decreases toward the edge of the accessible spectral range.

The incident power was varied from $100$ to $1000~\mu\mathrm{W}$ using a Gaussian excitation beam with a $1/e^{2}$ intensity radius of $w_0 = 15~\mu\mathrm{m}$ (\nameref{sec:Methods}), corresponding to intensities of approximately $50$--$300~\mathrm{W~cm^{-2}}$. The power sequence was randomized at each wavelength to minimize any hysteresis effects. The corresponding chopped-light chronoamperometry traces for the redox reaction are provided in Supporting Figure ~8. Figure~\ref{fig:secm_redox}e plots $\Delta I_{\mathrm{tip}}$ as a function of intensity for the three wavelengths. The largest photocurrent is observed at 760 nm, followed by 720 nm and then 800 nm, consistent with the resonance-enhanced response in Figure~\ref{fig:secm_redox}d.

The power-dependent responses at all three wavelengths are well described by linear fits, shown as dashed lines, with $R^2 > 0.995$ (Supporting Table 1(a)). Error bars represent the standard deviation from three repeated chopped-light cycles at each wavelength. 

The dependence of photocurrent on illumination intensity provides an additional diagnostic for distinguishing photon-driven and thermally assisted catalytic processes. In plasmonic photocatalysis, superlinear power-law scaling is often associated with thermal or hot-carrier-mediated enhancement, with reaction rates scaling as $R \propto I^n$ where $n > 1$ \cite{verma_paradox_2024,chen_origin_2023}. Sublinear behavior, often observed in semiconductor photocatalysts, can indicate carrier recombination, reactant mass-transport limitations, or temperature-dependent decreases in reaction rate \cite{chen_origin_2023,christopher_singular_2012,ghoshal_catalyst-free_2019,bell_light_2013}. Within the intensity window accessible in our setup, the metasurface response remains linear and does not transition to either sublinear or superlinear scaling. This supports assignment of the approximately tenfold increase in $\Delta I_{\mathrm{tip}}$ at the 760 nm resonance to photon-driven charge generation rather than photothermal acceleration.

We performed the same power-law analysis for the unpatterned thin film (Figure~\ref{fig:secm_redox}f). The photocurrent increased linearly with illumination intensity at all three wavelengths, with $R^2 > 0.995$ (Supporting Table~1(b)). The highest response was observed at 720~nm, followed by 760 and 800~nm, consistent with the wavelength-dependent optical absorption of the thin film. Because the thin-film absorptance is relatively similar across these wavelengths, the corresponding photocurrent slopes are also comparable. By contrast, the non-uniform absorptance spectrum of the metasurface produces a steeper slope at the 760~nm resonance, directly linking the wavelength-dependent photocurrent response to resonant optical absorption.

Consequently, we quantified the wavelength-dependent external quantum efficiency (EQE), also referred to as the incident-photon-to-current efficiency (IPCE), for the unpatterned thin film and the $w = 130~\mathrm{nm}$ metasurface (Figure~\ref{fig:secm_redox}g). The analysis procedure is described in Supporting Notes 1 and 2. Because photo-SECM measures the mediator current at the tip rather than the substrate photocurrent directly, the measured $\Delta I_{\mathrm{tip}}$ was converted into the wavelength-dependent substrate photocurrent, $I_{\mathrm{sub,photo}}(\lambda)$, using a finite-element diffusion model.

The tip-current response was simulated using a two-dimensional axisymmetric steady-state diffusion model in COMSOL Multiphysics, adapted from previously reported photo-SECM frameworks~\cite{sanchez-sanchez_scanning_2008,kiani_transport_2023}. The model accounts for diffusion between the substrate and tip, tip-generation/substrate-collection boundary conditions, and the Gaussian illumination profile at the substrate. The governing diffusion equation, boundary conditions, substrate reaction term, and current-integration procedure are described in Supporting Note 2 and Supporting Figure~9. The extracted $I_{\mathrm{sub,photo}}(\lambda)$ was divided by the incident photon flux to obtain the EQE.

The EQE spectra of the metasurface and thin film are comparable over much of the measured wavelength range, while the metasurface exhibits local enhancements near the resonant regions around 650 and 760~nm. These spectral features broadly follow the absorptance response, demonstrating that the engineered optical resonances are translated into enhanced external photocurrent generation. The correspondence is particularly pronounced near the guided-mode resonance at the a-Si band-edge, where the metasurface maintains an enhanced photoelectrochemical response despite the intrinsically weak absorption of the unpatterned film.

To evaluate the collected photocurrent per absorbed photon, we calculated the internal quantum efficiency (IQE) by normalizing the EQE by the experimentally measured absorptance (Figure~\ref{fig:secm_redox}h). Each photo-SECM measurement samples an approximately 10-nm illumination bandwidth; the absorptance used in the IQE calculation was therefore averaged over the corresponding $\pm 5$-nm interval. The propagated uncertainty includes contributions from the photocurrent and absorptance measurements. For the metasurface, the absorptance uncertainty also accounts for spatial variations across the measured region and an additional 10\% relative uncertainty.

The IQE analysis reveals a substantially larger difference between the two architectures near the a-Si band-edge, with the metasurface reaching up to approximately ten times the IQE of the unpatterned thin film. The IQE enhancement exceeds the corresponding EQE enhancement, showing that the resonant architecture increases not only the number of absorbed photons but also the fraction that contributes to interfacial charge transfer. The enhancement persists on the red side of the guided-mode resonance, where the metasurface absorptance decreases while the photoelectrochemical response remains comparatively strong, indicating efficient utilization of absorbed photons in this near-infrared spectral region.

To further probe the carrier dynamics underlying the different absorbed-photon conversion efficiencies of the thin film and metasurface, we performed femtosecond transient reflectivity measurements under above-gap excitation. Experimental details, including pump/probe conditions, fluence calibration, and fitting procedures, are described in the Supporting Note 3. The measurement configuration is shown schematically in Supporting Figure 11, and the full transient spectra, fluence-dependent traces, and fitting results are provided in Supporting Figure 12.  

A 390 nm pump photon energy, corresponding to 3.18 eV, was used to excite carriers well above the optical gap of the sputtered a-Si, determined by Tauc analysis to be 1.79 eV. The transient response was monitored near 372 nm, a spectral region chosen to minimize contributions from metasurface resonances and thin-film interference features. Because both the UV pump and probe are strongly absorbed in a-Si, the optical penetration depth is limited to approximately 12 nm, making the measurement primarily sensitive to near-surface carrier relaxation. This surface sensitivity is particularly relevant for comparison with the photoelectrochemical measurements, where interfacial charge transfer occurs at the semiconductor/electrolyte interface.

 Both the planar thin film and nanopatterned metasurface show an ultrafast reflectivity change followed by a long-lived plateau, consistent with rapid carrier trapping into localized defect states and a slower thermo-optical contribution \cite{esser_ultrafast_1990,della_valle_nonlinear_2017}. The initial component scales linearly with excitation density, while the characteristic trapping time remains nearly fluence-independent over the measured range, indicating that the measurements are performed below the high-density nonlinear recombination and below trap saturation regimes. The metasurface exhibits a similar relaxation pathway but with a shorter trapping time, consistent with additional near-surface trapping channels introduced by nanostructuring and etching.  These results support a picture in which in the case of the metasurface, carriers generated by resonant absorption near the chemically active surface are more effectively localized and participate in interfacial charge transfer under steady-state photoelectrochemical operation.

Altogether the photo-SECM measurements establish a direct link between nanophotonic resonance engineering and wavelength-selective chemical activity. The metasurface preserves a linear, photon-driven response over the measured power range, enhances photocurrent generation at the near-infrared resonance, and improves the absorbed-photon-to-current conversion efficiency near the a-Si band-edge.

\section{Photo-SECCM measurements for the hydrogen evolution reaction} 
\label{sec:seccm-her}
To determine whether the resonance-enhanced response observed by photo-SECM extends to solar-fuel chemistry, we employed wavelength-selective light-coupled scanning electrochemical cell microscopy (photo-SECCM) to probe the hydrogen evolution reaction (HER) on silicon-based photocathodes. By confining the electrolyte meniscus to form a local electrochemical cell, SECCM enables sensitive, direct measurements of structure--activity relationships in (photo)electrocatalysts \cite{aaronson_scanning_2015,wahab_scanning_2020}.
Measurements were performed under both photocatalytic (PC) and photoelectrochemical (PEC) conditions. Silicon-based photocathodes are particularly relevant for solar-driven water splitting because they combine strong visible-light absorption, material abundance, and compatibility with established microfabrication approaches \cite{boettcher_photoelectrochemical_2011}.

As in the SECM experiments, the samples were illuminated from the bottom side. A single-barrel glass nanopipette, with a representative SEM micrograph shown in Supporting Figure 10, was filled with 0.05~M H$_2$SO$_4$ (pH$\approx$1) and equipped with a quasi-reference counter electrode (QRCE). This electrolyte was selected because acidic conditions favor the HER and are commonly used for semiconductor-based photocathodes \cite{urbain_development_2014,boettcher_photoelectrochemical_2011,feng_hydrogen_2018,fan_high_2022}. The measurements were performed on the as-prepared, native-oxide-terminated a-Si surface, consistent with reports that SiO$_x$ can protect silicon against oxidative corrosion under acidic photoelectrochemical conditions \cite{anderson_silicon_2017,esposito_h2_2013}.

\begin{figure}[H]
\centering
\includegraphics[width=17cm]{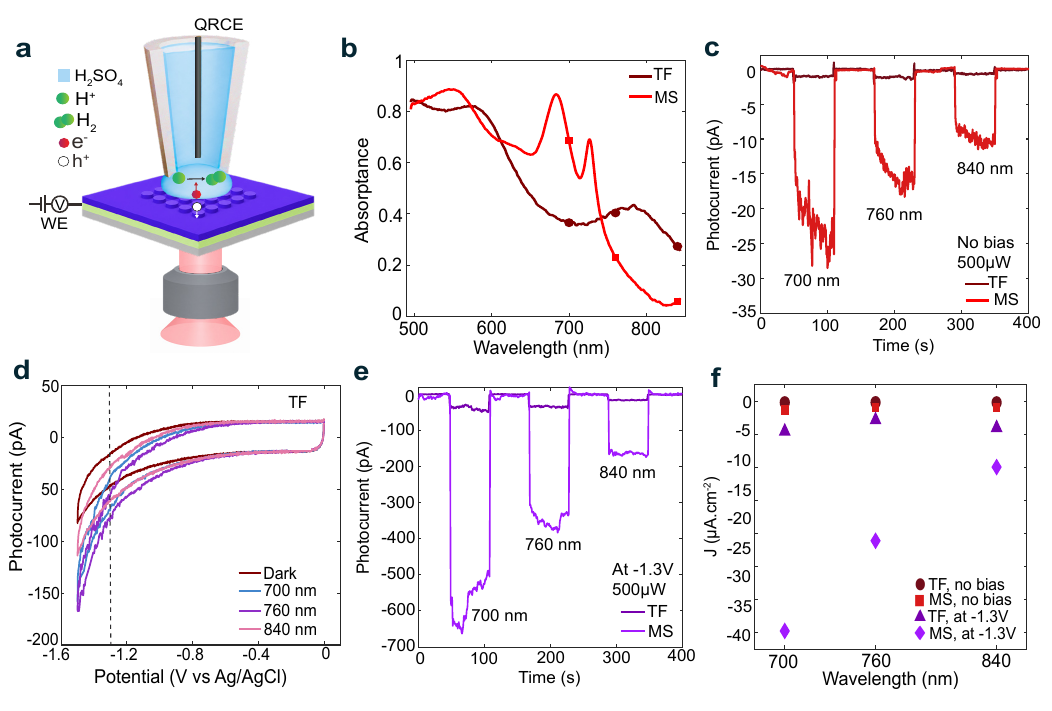}
\caption{\justifying{ Light-coupled scanning electrochemical cell microscopy (photo-SECCM) measurements toward hydrogen evolution reaction (HER).
\textbf{a} Schematic of the photo-SECCM configuration used to probe HER. The amorphous silicon (a-Si) thin film or metasurface serves as the working electrode (WE), while a quasi-reference counter electrode (QRCE) is inserted into the single-barrel glass nanopipette filled with 0.05 M H$_2$SO$_4$ electrolyte.
\textbf{b} Experimental absorptance spectra of the unpatterned thin film (TF) and the metasurface. For these measurements a different metasurface (MS) with \textit{w} = 210 and \(h_1=150~\mathrm{nm}\) was used. The wavelengths selected for chronoamperometric and cyclic voltammetric measurements, 700, 760 and 840 nm, are marked on the spectra.
\textbf{c}  Wavelength-dependent photocatalytic current measured for 0 V vs OCP for the TF and MS. The metasurface shows enhanced photocurrent relative to the TF, with the spectral trend following the absorptance enhancement in (b).
\textbf{d} Cyclic voltammograms measured on the TF under dark and illuminated conditions at 700, 760 and 840 nm, confirming the wavelength-dependent HER response. The dashed line indicates the bias values selected for subsequent photoelectrochemical measurements.
\textbf{e} Chronoamperometric response of the TF and MS at $-1.3$ V versus Ag/AgCl under wavelength-dependent illumination. The metasurface exhibits enhanced photoelectrochemical HER current compared with the TF while retaining a resonance-dependent spectral response.
\textbf{f} Wavelength-dependent photocurrent density, $J$, measured for the unpatterned thin film (TF) and metasurface (MS) at 0 V vs OCP and at $-1.3$ V vs Ag/AgCl.}}
\label{fig:seccm_her}
\end{figure}

Upon contact between the droplet and the sample, a confined electrochemical cell was formed at the illuminated surface, as schematically shown in Figure~\ref{fig:seccm_her}a. Figure~\ref{fig:seccm_her}b shows the absorptance spectra measured in air for the thin film and the metasurface. The metasurface used for these experiments is slightly different from the one previously discussed, with \textit{w} = 210, \(h_1=150~\mathrm{nm}\) and $p = 325$~nm. As discussed in the earlier section \nameref{sec:aSi metasurface}, immersion in electrolyte is expected to induce only a small spectral shift of up to approximately 15~nm. This shift should not alter the relative absorption trend at the selected excitation wavelengths. For the metasurface, absorption is expected to remain strongest at 700~nm, followed by 760~nm and then 840~nm, whereas the thin film shows an interference-related maximum near 760~nm.

Photocurrent transients within the micro-scale electrochemical cell were recorded at 700, 760, and 840~nm using chopped-light chronoamperometry with 30~s light-on and light-off intervals. The incident power was fixed at 500~$\mu$W, selected from the middle of the power range examined in the SECM measurements. Under PC conditions, the metasurface generated a cathodic photocurrent of up to approximately -25~pA at 700~nm, with smaller responses at 760 and 840~nm. The spike near 90~s and the current fluctuations are attributed to hydrogen bubble formation and release during the gas-forming reaction. This wavelength dependence follows the metasurface absorption trend in Figure~\ref{fig:seccm_her}b.

For the thin film, the photocurrents at 700 and 760~nm were comparable, with 760~nm giving a slightly larger response, whereas 840~nm produced the weakest photocurrent. This behavior is consistent with the thin-film absorptance spectrum, where the interference fringe between 700 and 800~nm gives stronger absorption near 760~nm. Across all three wavelengths, however, the metasurface produced larger photocurrents than the unpatterned thin film. At 0~V vs OCP, the metasurface response was enhanced by factors of $21.2 \pm 2.8$, $12.8 \pm 1.9$, and $14.6 \pm 1.8$ at 700, 760, and 840~nm, respectively.

To better resolve the modest wavelength-dependent differences observed for the thin film in the chronoamperometry traces, cyclic voltammograms were recorded on the thin film in the dark and under illumination at the three selected wavelengths. These measurements also allow the bias dependence of the thin-film photocurrent to be examined. As shown in Figure~\ref{fig:seccm_her}d, the cathodic current increased with an increasingly negative potential, confirming enhanced PEC HER activity under bias. The spectral trend followed the absorption response of the thin film, with the largest photocurrent at 760~nm, followed by 700~nm and then 840~nm. At -1.5~V vs Ag/AgCl, the photocurrent was approximately twice that measured at -1.3~V vs Ag/AgCl.

We then selected −1.3 V vs Ag/AgCl as a representative cathodic bias to compare the metasurface and thin film under photoelectrochemical conditions (Figure~4e). This potential produced a clear wavelength-dependent photoresponse, while more negative potentials led to increased gas evolution and a noisier SECCM response. Relative to the photocatalytic condition at 0 V vs OCP, applying a bias of −1.3 V increased the metasurface photocurrent by 543 pA at 700 nm, 328 pA at 760 nm, and 146 pA at 840 nm. Both samples exhibited increased photocurrent under bias, while the metasurface remained substantially more active at all three wavelengths. Relative to the thin film, the metasurface photocurrent was enhanced by factors of $14.70 \pm 2.51$, $10.24 \pm 0.67$, and $10.15 \pm 0.45$ at 700, 760, and 840~nm, respectively. Overall, the HER-associated photocurrent response is governed by both the resonant nanostructured geometry and the applied cathodic bias.

Figure~\ref{fig:seccm_her}f shows the photocurrent density normalized by the estimated electrochemically accessible area wetted by the \(33~\mu\mathrm{m}\)-diameter SECCM droplet. The thin-film area was \(8.55\times10^{-6}~\mathrm{cm}^{2}\). Accounting for the additional surface area introduced by the nanopillars gave an area-enhancement factor of 1.9 and a total metasurface area of \(1.6\times10^{-5}~\mathrm{cm}^{2}\). The current density was calculated as \(J=I_{\mathrm{photo}}/A_{\mathrm{real}}\), allowing the samples to be compared per unit of accessible surface area. Further details are provided in Supporting Note 4. Taking \(700~\mathrm{nm}\) as an example, the metasurface photocurrent was approximately 21-fold higher than that of the thin film at \(0~\mathrm{V}\) vs. OCP and 14.7-fold higher at \(-1.3~\mathrm{V}\) vs. Ag/AgCl. After surface-area normalization, \(\lvert J\rvert\) enhancements of 11.2-fold and 7.7-fold remained, respectively. The absorptance increased from approximately \(36\%\) for the thin film to \(68\%\) for the metasurface, corresponding to an optical enhancement of 1.9-fold. Assuming that photocurrent scales proportionally with absorption and electrochemically accessible surface area, these contributions account for an expected enhancement of approximately 3.6-fold. The larger experimental enhancement indicates that absorption and surface area alone are insufficient to explain the metasurface response, suggesting additional improvements in photogenerated-carrier separation, collection, or interfacial charge transfer enabled by the metasurface.

\section{Discussion} 
\label{sec:Discussion}

Optical metasurfaces offer a powerful route to overcome the limitations of planar photocathodes by enabling precise spectral and spatial control over light-driven chemical processes\cite{huttenhofer_metasurface_2021,paudel_metasurface-enhanced_2024,hu_catalytic_2022}. Although amorphous silicon (a-Si) is widely used in nanophotonics, its potential as an intrinsically active material for photocatalysis and photoelectrochemistry remains comparatively underexplored. Here, we demonstrate an all-dielectric a-Si metasurface engineered to couple optical resonances directly to interfacial chemical reactivity. The metasurface achieves absorptance above 80\% near the a-Si band-edge, compared with less than 30\% for the unpatterned thin film, while simultaneously functioning as the light absorber, charge-transport medium, and catalytic surface without an intentionally deposited co-catalyst or passivation layer.

Wavelength- and power-resolved photo-SECM measurements establish a direct correspondence between the resonant optical response and local redox activity. The EQE spectra reproduce the resonance-associated spectral features, showing that optical confinement is translated into enhanced external photocurrent generation. After normalization by the experimentally measured absorptance, the metasurface reaches up to approximately ten times the IQE of the unpatterned film near the a-Si band-edge, demonstrating more effective conversion of absorbed photons into interfacial current. Photo-SECCM measurements further reveal enhancements of up to 21-fold under photocatalytic conditions and 15-fold under photoelectrochemical conditions relative to the planar thin film. Even after normalization by the estimated wetted surface area, current-density enhancements of 11.2-fold at 0~V versus OCP and 7.7-fold at $-1.3$~V versus Ag/AgCl remain at 700~nm, exceeding the corresponding 1.9-fold absorptance enhancement. These results show that the improved activity cannot be explained by enhanced absorption and increased surface area alone, but is also consistent with more effective utilization of photogenerated carriers for interfacial charge transfer.

The metasurface retains its resonant optical response and wavelength-dependent photocurrent during prolonged aqueous operation, while post-reaction characterization confirms preservation of the nanopillar morphology and multilayer architecture. Beyond demonstrating enhanced photoelectrochemical performance, this work uses wavelength-resolved photo-SECM and photo-SECCM to interrogate the semiconductor locally under operando conditions and directly connect resonant absorption, absorbed-photon conversion, and interfacial chemical activity. The combined EQE and IQE analysis distinguishes enhanced light harvesting from improved utilization of absorbed photons, while surface-area-normalized measurements show that the increased reactivity cannot be attributed to geometry alone. These results establish resonantly engineered a-Si metasurfaces as stable, monolithic photoelectrodes for resolving and controlling light-driven chemistry through nanoscale geometry. Their compatibility with established thin-film deposition and semiconductor-processing technologies provides a route toward scalable solar-fuel devices and integrated wavelength-selective photochemical systems\cite{li_photoelectrochemical_2019,mascaretti_designing_2023,hu_catalytic_2022}.

\section{Methods} \label{sec:Methods}
\subsection*{Numerical Simulations}

Full-wave electromagnetic simulations were performed using the Wave Optics Module in COMSOL Multiphysics. The model consisted of a three-dimensional unit cell comprising an amorphous silicon (a-Si) meta-atom on an indium tin oxide (ITO) layer supported by a fused silica substrate. Simulations were conducted under bottom illumination in both air and water to replicate the experimental conditions of the photo-SECM measurements. Wavelength-dependent optical constants of a-Si and ITO were experimentally determined by spectroscopic ellipsometry (Woollam RC2). The data were fitted in CompleteEASE using a Cody–Lorentz oscillator model with an effective surface-roughness layer for a-Si and a Drude–Lorentz model for ITO. The extracted refractive indices and extinction coefficients were used as inputs to the electromagnetic simulations.

Floquet-periodic boundary conditions were applied in the lateral directions to represent an infinite array, while perfectly matched layers were implemented along the propagation direction to absorb outgoing radiation and suppress spurious reflections. The structure was excited through a periodic port using a normally incident plane wave over the wavelength range 450--850~nm. Reflectance and transmittance were obtained from the outgoing power at the input and output ports, respectively, and absorptance was calculated as $A=1-R-T$. To accurately account for guided-mode resonances, the periodic ports included diffraction orders $m,n=-2,-1,0,+1,+2$ along the two in-plane lattice directions.

\subsection*{Nanofabrication of meta-electrodes}
Metasurfaces were fabricated on 520 $\mu$m thick 2 × 2 cm$^2$ fused silica substrates, as illustrated in Supporting Figure 1. A 30 nm indium tin oxide (ITO) layer was deposited onto clean substrates by RF sputtering (Pfeiffer SPIDER 600) to serve as a transparent conductive electrode and hole-transport layer. Then, a nominally 220-nm-thick amorphous silicon (a-Si) layer was deposited at room temperature (Alliance-Concept DP 650), leaving approximately one quarter of the ITO exposed for electrical contact. Nanodisk arrays of 55 x 55 $\mu\mathrm{m}^2$ or 110 x 110 $\mu\mathrm{m}^2$ were patterned on a-Si by spin-coating a 120 nm layer of ZEP 520A electron-beam resist, followed by an Electra 92 conductive polymer overlayer (Zeon Chemicals) to mitigate charging during exposure. Electron-beam lithography was subsequently performed using a Raith EBPG5000+ operated at 100 kV with exposure doses ranging from 140 to 240 $\mu$C cm$^{-2}$.
After development, the resist served as an etch mask for pattern transfer into the a-Si film. Exposed a-Si was removed by anisotropic reactive ion etching (Adixen AMS200, 0 $^\circ$C), yielding nanodisks with near-vertical sidewalls. Residual resist was removed by acetone immersion, followed by microwave oxygen plasma cleaning (Tepla 300, 500 W, 400 mL min$^{-1}$ O$_2$, 30 s).

\subsection*{Characterization of meta-electrodes}
\label{sec:characterization}
The metasurfaces were characterized by scanning electron microscopy (SEM; Zeiss Crossbeam and Merlin) and atomic force microscopy (AFM; Bruker FastScan). The wavelength-dependent refractive index and extinction coefficient of the amorphous silicon and ITO layers were determined by spectroscopic ellipsometry (Woollam RC2) and used as inputs for all electromagnetic simulations (see Supporting Figure 3).

Microscale absorptance measurements were carried out on a NT\&C NanoMicroSpec platform consisting of an optical microscope (Nikon Eclipse Ti2) coupled to a spectrometer (Princeton Instruments SpectraPro HRS-500) and illuminating the sample from the bottom by a laser-driven white-light source (Energetica LDLS) delivered through a 100 $\mu$m core optical fiber. For transmission measurements, the incident beam was collimated using a condenser and referenced to a bare fused silica substrate. For reflection measurements, the beam was focused at the center of the objective back focal plane using a $60\times$ objective (Nikon S Plan Fluor ELWD, NA = 0.7). Reflectance was calibrated against a protected silver mirror with known spectral response (Thorlabs PF10-03-P01).

Cross-sectional transmission electron microscopy (TEM) specimens were prepared using a dual-beam focused ion beam/scanning electron microscope (FIB/SEM; Zeiss NVision 40), using a methodology previously described in \cite{dayi_large-area_2025}. The region of interest was first protected with an approximately 1 $\mu$m thick carbon layer deposited by ion-beam-assisted deposition. Lamellae were then milled using a 30 kV Ga$^+$ ion beam, lifted out, and transferred onto Cu TEM grids.

The amorphous nature of the a-Si film was confirmed by X-ray diffraction (XRD) measurements, shown in the smoothed XRD pattern in Supporting Figure 2, which exhibits a broad diffuse hump without well-defined Bragg diffraction peaks. The XRD measurement was performed in a D8 DISCOVER Plus diffractometer in grazing angle diffraction configuration.

\subsection*{Photo-SECM measurements for Redox Couple Reactions}

 Photoelectrochemical measurements were performed using a custom-built photo-SECM platform previously described in detail by Kiani \textit{et al.}\cite{kiani_transport_2023}. Briefly, the setup comprised an inverted optical microscope (Nikon Eclipse Ti2), a custom electrochemical cell, a bipotentiostat (Biologic SP-300), and a Pt ultramicroelectrode (UME) mounted on a piezoelectric positioning stage (MMP1/NanoF450, Mad City Labs). A conventional three-electrode configuration was employed, with the a-Si metasurface serving as the working electrode, a Pt wire (0.5 mm diameter) as the counter electrode, and an Ag/AgCl reference electrode (LF1, Innovative Instruments). Unless otherwise stated, measurements were carried out in 0.2 M aqueous sodium sulfate containing 2 mM potassium ferrocyanide. Electrical contact was made through an exposed region of the  ITO that was isolated from the electrolyte.

A supercontinuum white-light laser (NKT Photonics), coupled to a tunable wavelength filter (SuperK VARIA), was used to control the excitation wavelength and power. The excitation beam was modulated using an optical shutter (Thorlabs SH1) and focused in a bottom-illumination configuration, producing a Gaussian beam with a 1/$e^2$ radius of approximately 15 $\mu$m, as determined by CCD beam profiling. The ferro/ferricyanide redox couple does not absorb over the investigated spectral range, therefore photogenerated charge carriers originate exclusively from the a-Si photoelectrode \cite{kiani_transport_2023} .

Pt and glass ultramicroelectrodes (UMEs) were fabricated following a previously described laser-pulling method \cite{mezour_fabrication_2011}. During the measurements, the UME of 13.5$\mu$m diameter was positioned approximately 10 $\mu$m above the sample surface using the piezoelectric stage. At each wavelength, the tip photocurrent was measured using chopped illumination with three light-on/light-off cycles, each with a 3 s illumination period. The initial transient following illumination was excluded from the analysis (see Supporting Figure 8). The photocurrent was calculated using a custom analysis script that identified the illumination cycles, excluded the transient response, and determined the average tip photocurrent and standard deviation from the three light-on periods.

\subsection*{Photo-SECCM measurements for the hydrogen evolution reaction}

Photo-SECCM measurements were performed using a single-barrel glass nanopipette to locally probe the hydrogen evolution reaction (HER) on the a-Si thin film and metasurface photoelectrodes. A representative SEM image of the nanopipette is shown in Supporting Figure 10. The nanopipette was filled with 0.05 M H$_2$SO$_4$ (pH 1), a commonly used electrolyte for evaluating silicon-based photocathodes for hydrogen evolution, and contained a quasi-reference counter electrode (QRCE). Upon contact between the nanopipette meniscus and the sample surface, a confined electrochemical cell was formed locally on the illuminated photoelectrode.

Samples were illuminated from the substrate side using the same supercontinuum laser source and wavelength-selection optics described for the photo-SECM measurements. Wavelength-dependent measurements were performed at 700, 760, and 840 nm using chopped illumination with alternating 30 s light-on and light-off intervals. Unless otherwise stated, the incident optical power was fixed at 500 $\mu$W. Chronoamperometric measurements were performed at 0 V vs OCP and under an applied potential of $-1.3$ V versus Ag/AgCl to evaluate the photocurrent associated with photocatalytic and photoelectrochemical hydrogen evolution, respectively. Cyclic voltammetry measurements were additionally performed to characterize the potential dependence of the HER response under dark and illuminated conditions.

\subsection*{Data Analysis and Figure Preparation}

Experimental data were processed and analyzed using custom MATLAB and Python scripts. Image analysis was performed using ImageJ and figures were assembled using Adobe Illustrator. OpenAI ChatGPT (GPT-5.6 Thinking) was used for language editing and coding assistance. The Table of Contents graphic was prepared using the image-generation tool integrated into ChatGPT, based on sketches and scientific direction provided by the authors. All AI-assisted outputs were reviewed and verified by the authors and AI tools were not used to generate or interpret experimental data.
 \newpage

%% file: sections/data.tex
\section*{Data Availability Statement} \label{sec:data}
Data supporting the findings of this study will be available at Zenodo.

%% file: sections/supportinginfo.tex
\section*{Supporting Information} \label{sec:supportinginfo} 

Supporting Information is available for this paper.

%% file: sections/acknowledgements.tex
\section*{Acknowledgements} \label{sec:acknowledgements}
 E.N.D., P.V.  and  G.T. acknowledge the SNSF Eccellenza  Grant PCEGP2-194181. The authors also acknowledge the support of the following experimental facilities at EPFL: Centre for MicroNanofabrication (CMi),  EPFL: Interdisciplinary Centre for Electron Microscopy (CIME).  We would like to acknowledge Dr. Narmada Naidu Gopal for assistance in setting up the COMSOL models. We thank Dr. Zdenek Benes for valuable insights on optimizing the e-beam lithography fabrication procedure. We thank Dr. Fatemeh Kiani for her contributions to the photo-SECM setup and COMSOL Diffusion Model, and German Garcia Martinez for the software development for the SECM measurements. Finally, the authors would like to thank Dr. Lucie Navratilova for TEM lamella preparation with FIB-SEM and Dr. David Reyes for the TEM data acquisition. 

%% file: sections/contributions.tex
\section*{Author contributions} \label{sec:contributions}
G.T. and E.N.D. conceptualized the project. E.N.D. designed, fabricated, and characterized the metasurfaces. E.N.D. and P.V. carried out the SECM measurements. O.C.K. contributed to the absorption measurements. E.N.D. and M.S. performed SECCM measurements and COMSOL models for IQE calculation. M.S. contributed to discussions on the SECM setup. M.P. performed and analyzed transient reflectivity measurements with help from E.N.D.. E.N.D. analyzed the data and wrote the manuscript with input from all authors. G.T. supervised all aspects of the project.

%% file: sections/comp.tex
\section*{Competing interests} \label{sec:comp}
The Authors declare no competing interests.

%% file: sections/toc.tex
\section*{Table of Content Graphic} \label{sec:toc}

\begin{figure}[ht]
\centering
\includegraphics[width=0.85\textwidth]{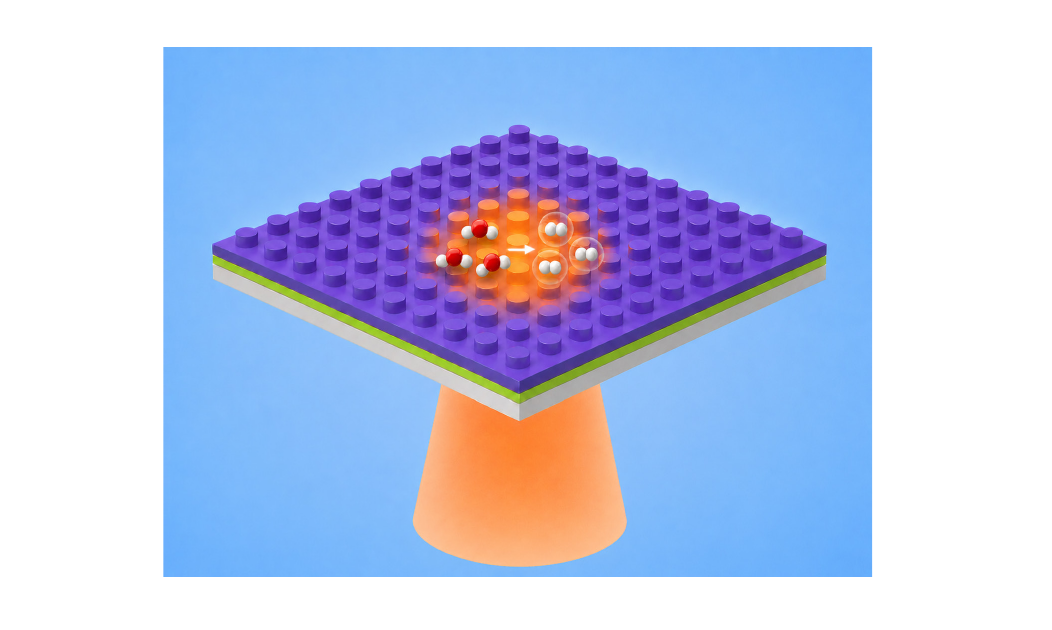}
\caption*{\small
An all-dielectric amorphous-silicon metasurface couples tunable optical resonances directly to interfacial chemistry within a 220-nm-thick photoelectrode. Resonant light confinement enhances near-band-edge absorption and local redox and hydrogen-evolution activity without added co-catalysts or engineered passivation. Spatially resolved electrochemical measurements reveal how metasurface geometry encodes wavelength-selective photochemical reactivity.
}
\end{figure}

%% file: ref.bib
@article{karaman_decoupling_2025,
	title = {Decoupling {Optical} and {Thermal} {Dynamics} in {Dielectric} {Metasurfaces} for {Self}‐{Encoded} {Photonic} {Control}},
	volume = {19},
	issn = {1863-8880, 1863-8899},
	url = {https://onlinelibrary.wiley.com/doi/10.1002/lpor.202501014},
	doi = {10.1002/lpor.202501014},
	language = {en},
	number = {24},
	urldate = {2026-07-26},
	journal = {Laser \& Photonics Reviews},
	author = {Karaman, Omer Can and Naidu, Gopal Narmada and Bowman, Alan R. and Dayi, Elif Nur and Tagliabue, Giulia},
	month = dec,
	year = {2025},
	pages = {2501014},
}

@article{dayi_large-area_2025,
	title = {Large-{Area} {Monocrystalline} {Copper} {Microflake} {Synthesis}},
	volume = {129},
	copyright = {https://creativecommons.org/licenses/by/4.0/},
	issn = {1932-7447, 1932-7455},
	url = {https://pubs.acs.org/doi/10.1021/acs.jpcc.5c00654},
	doi = {10.1021/acs.jpcc.5c00654},
	language = {en},
	number = {25},
	urldate = {2026-07-14},
	journal = {The Journal of Physical Chemistry C},
	author = {Dayi, Elif Nur and Pellet, Diotime and Vensaus, Priscila and Kiani, Fatemeh and Bowman, Alan R. and Karaman, Omer Can and Tagliabue, Giulia},
	month = jun,
	year = {2025},
	pages = {11574--11582},
}

@article{della_valle_nonlinear_2017,
	title = {Nonlinear {Anisotropic} {Dielectric} {Metasurfaces} for {Ultrafast} {Nanophotonics}},
	volume = {4},
	url = {https://doi.org/10.1021/acsphotonics.7b00544},
	doi = {10.1021/acsphotonics.7b00544},
	number = {9},
	urldate = {2026-05-10},
	journal = {ACS Photonics},
	publisher = {American Chemical Society},
	author = {Della Valle, Giuseppe and Hopkins, Ben and Ganzer, Lucia and Stoll, Tatjana and Rahmani, Mohsen and Longhi, Stefano and Kivshar, Yuri S. and De Angelis, Costantino and Neshev, Dragomir N. and Cerullo, Giulio},
	month = sep,
	year = {2017},
	pages = {2129--2136},
}

@article{esposito_h2_2013,
	title = {H2 evolution at {Si}-based metal–insulator–semiconductor photoelectrodes enhanced by inversion channel charge collection and {H} spillover},
	volume = {12},
	issn = {1476-1122, 1476-4660},
	url = {https://www.nature.com/articles/nmat3626},
	doi = {10.1038/nmat3626},
	language = {en},
	number = {6},
	urldate = {2026-08-03},
	journal = {Nature Materials},
	author = {Esposito, Daniel V. and Levin, Igor and Moffat, Thomas P. and Talin, A. Alec},
	month = jun,
	year = {2013},
	pages = {562--568},
}

@article{anderson_silicon_2017,
	title = {Silicon {Photoelectrode} {Thermodynamics} and {Hydrogen} {Evolution} {Kinetics} {Measured} by {Intensity}-{Modulated} {High}-{Frequency} {Resistivity} {Impedance} {Spectroscopy}},
	volume = {8},
	issn = {1948-7185},
	url = {https://doi.org/10.1021/acs.jpclett.7b01311},
	doi = {10.1021/acs.jpclett.7b01311},
	number = {21},
	urldate = {2026-08-03},
	journal = {The Journal of Physical Chemistry Letters},
	author = {Anderson, Nicholas C. and Carroll, Gerard M. and Pekarek, Ryan T. and Christensen, Steven T. and van de Lagemaat, Jao and Neale, Nathan R.},
	month = oct,
	year = {2017},
	pages = {5253--5258},
}

@article{urbain_development_2014,
	title = {Development of {Thin} {Film} {Amorphous} {Silicon} {Tandem} {Junction} {Based} {Photocathodes} {Providing} {High} {Open}-{Circuit} {Voltages} for {Hydrogen} {Production}},
	volume = {2014},
	copyright = {Copyright © 2014 F. Urbain et al.},
	issn = {1687-529X},
	url = {https://onlinelibrary.wiley.com/doi/abs/10.1155/2014/249317},
	doi = {10.1155/2014/249317},
	language = {en},
	number = {1},
	urldate = {2026-08-03},
	journal = {International Journal of Photoenergy},
	author = {Urbain, F. and Wilken, K. and Smirnov, V. and Astakhov, O. and Lambertz, A. and Becker, J.-P. and Rau, U. and Ziegler, J. and Kaiser, B. and Jaegermann, W. and Finger, F.},
	year = {2014},
	note = {\_eprint: https://onlinelibrary.wiley.com/doi/pdf/10.1155/2014/249317},
	pages = {249317},
}

@article{li_photoelectrochemical_2019,
	title = {Photoelectrochemical {CO2} reduction to adjustable syngas on grain-boundary-mediated a-{Si}/{TiO2}/{Au} photocathodes with low onset potentials},
	volume = {12},
	issn = {1754-5706},
	url = {https://pubs.rsc.org/en/content/articlelanding/2019/ee/c8ee02768d},
	doi = {10.1039/C8EE02768D},
	language = {en},
	number = {3},
	urldate = {2026-03-20},
	journal = {Energy \& Environmental Science},
	publisher = {The Royal Society of Chemistry},
	author = {Li, Chengcheng and Wang, Tuo and Liu, Bin and Chen, Mengxin and Li, Ang and Zhang, Gong and Du, Minyong and Wang, Hui and Liu, Shengzhong Frank and Gong, Jinlong},
	month = mar,
	year = {2019},
	pages = {923--928},
}

@article{mezour_fabrication_2011,
	title = {Fabrication and {Characterization} of {Laser} {Pulled} {Platinum} {Microelectrodes} with {Controlled} {Geometry}},
	volume = {83},
	issn = {0003-2700},
	url = {https://doi.org/10.1021/ac102482f},
	doi = {10.1021/ac102482f},
	number = {6},
	urldate = {2026-07-28},
	journal = {Analytical Chemistry},
	author = {Mezour, Mohamed A. and Morin, Mario and Mauzeroll, Janine},
	month = feb,
	year = {2011},
	pages = {2378--2382},
}

@article{lin_amorphous_2013,
	title = {Amorphous {Si} {Thin} {Film} {Based} {Photocathodes} with {High} {Photovoltage} for {Efficient} {Hydrogen} {Production}},
	volume = {13},
	issn = {1530-6984},
	url = {https://doi.org/10.1021/nl403265k},
	doi = {10.1021/nl403265k},
	number = {11},
	urldate = {2026-07-27},
	journal = {Nano Letters},
	author = {Lin, Yongjing and Battaglia, Corsin and Boccard, Mathieu and Hettick, Mark and Yu, Zhibin and Ballif, Christophe and Ager, Joel W. and Javey, Ali},
	month = sep,
	year = {2013},
	pages = {5615--5618},
}

@article{tauc_optical_1966,
	title = {Optical {Properties} and {Electronic} {Structure} of {Amorphous} {Germanium}},
	volume = {15},
	copyright = {Copyright © 1966 WILEY-VCH Verlag GmbH \& Co. KGaA, Weinheim},
	issn = {1521-3951},
	url = {https://onlinelibrary.wiley.com/doi/abs/10.1002/pssb.19660150224},
	doi = {10.1002/pssb.19660150224},
	language = {en},
	number = {2},
	urldate = {2026-07-25},
	journal = {physica status solidi (b)},
	author = {Tauc, J. and Grigorovici, R. and Vancu, A.},
	year = {1966},
	note = {\_eprint: https://onlinelibrary.wiley.com/doi/pdf/10.1002/pssb.19660150224},
	pages = {627--637},
}

@article{cody_disorder_1981,
	title = {Disorder and the {Optical}-{Absorption} {Edge} of {Hydrogenated} {Amorphous} {Silicon}},
	volume = {47},
	copyright = {http://link.aps.org/licenses/aps-default-license},
	issn = {0031-9007},
	url = {https://link.aps.org/doi/10.1103/PhysRevLett.47.1480},
	doi = {10.1103/PhysRevLett.47.1480},
	language = {en},
	number = {20},
	urldate = {2026-07-25},
	journal = {Physical Review Letters},
	author = {Cody, G. D. and Tiedje, T. and Abeles, B. and Brooks, B. and Goldstein, Y.},
	month = nov,
	year = {1981},
	pages = {1480--1483},
}

@article{yang_optofluidic_2023,
	title = {Optofluidic transport and assembly of nanoparticles using an all-dielectric quasi-{BIC} metasurface},
	volume = {12},
	copyright = {2023 The Author(s)},
	issn = {2047-7538},
	url = {https://www.nature.com/articles/s41377-023-01212-4},
	doi = {10.1038/s41377-023-01212-4},
	language = {en},
	number = {1},
	urldate = {2026-07-22},
	journal = {Light: Science \& Applications},
	publisher = {Nature Publishing Group},
	author = {Yang, Sen and Ndukaife, Justus C.},
	month = jul,
	year = {2023},
	pages = {188},
}

@article{zhou_metasurface-assisted_2025,
	title = {Metasurface-assisted multimodal quantum imaging},
	volume = {122},
	url = {https://www.pnas.org/doi/abs/10.1073/pnas.2500760122},
	doi = {10.1073/pnas.2500760122},
	language = {en},
	number = {18},
	urldate = {2026-07-22},
	journal = {Proceedings of the National Academy of Sciences},
	publisher = {Proceedings of the National Academy of Sciences},
	author = {Zhou, Yifan and Zhu, Xiaoshu and Li, Tianyue and Zhou, Zhou and Bi, Qianhui and Liu, Jun and Chen, Jian and Fu, Boyan and He, Juanzi and Feng, Xiaojing and Feng, Xinyang and Liu, Xingyu and Wang, Qianjin and Wang, Shuming and Wang, Zhenlin and Qiu, Cheng-Wei and Zhu, Shining},
	month = may,
	year = {2025},
	pages = {e2500760122},
}

@article{chen_progress_2017,
	title = {Progress in {Tandem} {Solar} {Cells} {Based} on {Hybrid} {Organic}–{Inorganic} {Perovskites}},
	volume = {7},
	copyright = {© 2017 WILEY-VCH Verlag GmbH \& Co. KGaA, Weinheim},
	issn = {1614-6840},
	url = {https://onlinelibrary.wiley.com/doi/abs/10.1002/aenm.201602400},
	doi = {10.1002/aenm.201602400},
	language = {en},
	number = {14},
	urldate = {2026-07-09},
	journal = {Advanced Energy Materials},
	author = {Chen, Bo and Zheng, Xiaopeng and Bai, Yang and Padture, Nitin P. and Huang, Jinsong},
	year = {2017},
	note = {\_eprint: https://advanced.onlinelibrary.wiley.com/doi/pdf/10.1002/aenm.201602400},
	pages = {1602400},
}

@article{polman_photonic_2012,
	title = {Photonic design principles for ultrahigh-efficiency photovoltaics},
	volume = {11},
	copyright = {2012 Springer Nature Limited},
	issn = {1476-4660},
	url = {https://www.nature.com/articles/nmat3263},
	doi = {10.1038/nmat3263},
	language = {en},
	number = {3},
	urldate = {2026-07-08},
	journal = {Nature Materials},
	publisher = {Nature Publishing Group},
	author = {Polman, Albert and Atwater, Harry A.},
	month = mar,
	year = {2012},
	pages = {174--177},
}

@article{sanchez-sanchez_scanning_2008,
	title = {Scanning {Electrochemical} {Microscopy}. 60. {Quantitative} {Calibration} of the {SECM} {Substrate} {Generation}/{Tip} {Collection} {Mode} and {Its} {Use} for the {Study} of the {Oxygen} {Reduction} {Mechanism}},
	volume = {80},
	issn = {0003-2700, 1520-6882},
	url = {https://pubs.acs.org/doi/10.1021/ac702453n},
	doi = {10.1021/ac702453n},
	language = {en},
	number = {9},
	urldate = {2026-07-08},
	journal = {Analytical Chemistry},
	author = {Sánchez-Sánchez, Carlos M. and Rodríguez-López, Joaquín and Bard, Allen J.},
	month = may,
	year = {2008},
	pages = {3254--3260},
}

@article{ma_fundamental_2021,
	title = {Fundamental {Insights} into {Surface} {Modification} of {Silicon} {Material} toward {Improved} {Activity} and {Durability} in {Photocatalytic} {Hydrogen} {Production}: {A} {Case} {Study} of {Pre}-{Lithiation}},
	volume = {125},
	issn = {1932-7447},
	shorttitle = {Fundamental {Insights} into {Surface} {Modification} of {Silicon} {Material} toward {Improved} {Activity} and {Durability} in {Photocatalytic} {Hydrogen} {Production}},
	url = {https://doi.org/10.1021/acs.jpcc.1c00700},
	doi = {10.1021/acs.jpcc.1c00700},
	number = {10},
	urldate = {2026-07-06},
	journal = {The Journal of Physical Chemistry C},
	publisher = {American Chemical Society},
	author = {Ma, Jun and Gao, Chao and Low, Jingxiang and Liu, Dong and Lian, Xin and Zhang, Haochuan and Jin, Hongchang and Zheng, Xusheng and Wang, Chengming and Long, Ran and Ji, Hengxing and Zhu, Junfa and Xiong, Yujie},
	month = mar,
	year = {2021},
	pages = {5542--5548},
}

@article{huttenhofer_anapole_2020,
	title = {Anapole {Excitations} in {Oxygen}-{Vacancy}-{Rich} {TiO2}–x {Nanoresonators}: {Tuning} the {Absorption} for {Photocatalysis} in the {Visible} {Spectrum}},
	volume = {14},
	issn = {1936-0851},
	shorttitle = {Anapole {Excitations} in {Oxygen}-{Vacancy}-{Rich} {TiO2}–x {Nanoresonators}},
	url = {https://doi.org/10.1021/acsnano.9b09987},
	doi = {10.1021/acsnano.9b09987},
	number = {2},
	urldate = {2026-07-06},
	journal = {ACS Nano},
	publisher = {American Chemical Society},
	author = {Hüttenhofer, Ludwig and Eckmann, Felix and Lauri, Alberto and Cambiasso, Javier and Pensa, Evangelina and Li, Yi and Cortés, Emiliano and Sharp, Ian D. and Maier, Stefan A.},
	month = feb,
	year = {2020},
	pages = {2456--2464},
}

@article{yuan_quasi-bound_2024,
	title = {A {Quasi}-{Bound} {States} in the {Continuum} {Dielectric} {Metasurface}-{Based} {Antenna}–{Reactor} {Photocatalyst}},
	volume = {24},
	issn = {1530-6984},
	url = {https://doi.org/10.1021/acs.nanolett.3c03585},
	doi = {10.1021/acs.nanolett.3c03585},
	number = {1},
	urldate = {2026-07-06},
	journal = {Nano Letters},
	publisher = {American Chemical Society},
	author = {Yuan, Lin and Zhao, Yage and Toma, Andrea and Aglieri, Vincenzo and Gerislioglu, Burak and Yuan, Yigao and Lou, Minghe and Ogundare, Adebola and Alabastri, Alessandro and Nordlander, Peter and Halas, Naomi J.},
	month = jan,
	year = {2024},
	pages = {172--179},
}

@article{zhou_breaking_2025,
	title = {Breaking the photoelectrochemical activity-battery voltage trade-off for efficient photocharging of {TEMPO}/quinone redox flow battery},
	volume = {507},
	issn = {1385-8947},
	url = {https://www.sciencedirect.com/science/article/pii/S1385894725009672},
	doi = {10.1016/j.cej.2025.160162},
	urldate = {2026-07-06},
	journal = {Chemical Engineering Journal},
	author = {Zhou, Weicheng and Liu, Mingyao and Cao, Yuexian and Sun, Youming and Chang, Qingbo and Zhao, Xuefei and Wang, Hui and Liu, Shengzhong and Shi, Jingying and Li, Can},
	month = mar,
	year = {2025},
	pages = {160162},
}

@article{hu_catalytic_2022,
	title = {Catalytic {Metasurfaces} {Empowered} by {Bound} {States} in the {Continuum}},
	volume = {16},
	copyright = {https://creativecommons.org/licenses/by-nc-nd/4.0/},
	issn = {1936-0851, 1936-086X},
	url = {https://pubs.acs.org/doi/10.1021/acsnano.2c05680},
	doi = {10.1021/acsnano.2c05680},
	language = {en},
	number = {8},
	urldate = {2026-07-06},
	journal = {ACS Nano},
	author = {Hu, Haiyang and Weber, Thomas and Bienek, Oliver and Wester, Alwin and Hüttenhofer, Ludwig and Sharp, Ian D. and Maier, Stefan A. and Tittl, Andreas and Cortés, Emiliano},
	month = aug,
	year = {2022},
	pages = {13057--13068},
}

@article{wu_tio2_2019,
	title = {{TiO2} metasurfaces: {From} visible planar photonics to photochemistry},
	volume = {5},
	shorttitle = {{TiO2} metasurfaces},
	url = {https://www.science.org/doi/10.1126/sciadv.aax0939},
	doi = {10.1126/sciadv.aax0939},
	number = {11},
	urldate = {2026-07-06},
	journal = {Science Advances},
	publisher = {American Association for the Advancement of Science},
	author = {Wu, Yunkai and Yang, Wenhong and Fan, Yubin and Song, Qinghai and Xiao, Shumin},
	month = nov,
	year = {2019},
	pages = {eaax0939},
}

@article{mascaretti_designing_2023,
	title = {Designing {Metasurfaces} for {Efficient} {Solar} {Energy} {Conversion}},
	volume = {10},
	copyright = {https://creativecommons.org/licenses/by-nc-nd/4.0/},
	issn = {2330-4022, 2330-4022},
	url = {https://pubs.acs.org/doi/10.1021/acsphotonics.3c01013},
	doi = {10.1021/acsphotonics.3c01013},
	language = {en},
	number = {12},
	urldate = {2026-07-03},
	journal = {ACS Photonics},
	author = {Mascaretti, Luca and Chen, Yuheng and Henrotte, Olivier and Yesilyurt, Omer and Shalaev, Vladimir M. and Naldoni, Alberto and Boltasseva, Alexandra},
	month = dec,
	year = {2023},
	pages = {4079--4103},
}

@article{feng_hydrogen_2018,
	title = {Hydrogen evolution from silicon nanowire surfaces},
	volume = {8},
	issn = {2046-2069},
	url = {https://doi.org/10.1039/c8ra07905f},
	doi = {10.1039/c8ra07905f},
	number = {72},
	urldate = {2026-07-03},
	journal = {RSC Advances},
	author = {Feng, Rui and Liu, Yang and Li, Shipu and Chen, Hanbin and Song, Chengyi and Tao, Peng and Wu, Jianbo and Zhang, Peng and Deng, Tao and Shang, Wen},
	month = dec,
	year = {2018},
	pages = {41657--41662},
}

@article{fan_high_2022,
	title = {High {Entropy} {Alloy} {Electrocatalytic} {Electrode} toward {Alkaline} {Glycerol} {Valorization} {Coupling} with {Acidic} {Hydrogen} {Production}},
	volume = {144},
	issn = {0002-7863},
	url = {https://doi.org/10.1021/jacs.1c13740},
	doi = {10.1021/jacs.1c13740},
	number = {16},
	urldate = {2026-07-03},
	journal = {Journal of the American Chemical Society},
	publisher = {American Chemical Society},
	author = {Fan, Linfeng and Ji, Yaxin and Wang, Genxiang and Chen, Junxiang and Chen, Kai and Liu, Xi and Wen, Zhenhai},
	month = apr,
	year = {2022},
	pages = {7224--7235},
}

@article{wahab_scanning_2020,
	series = {Environmental {Electrochemistry} ● {Physical} and {Nano} {Electrochemistry}},
	title = {Scanning electrochemical cell microscopy: {A} natural technique for single entity electrochemistry},
	volume = {22},
	issn = {2451-9103},
	shorttitle = {Scanning electrochemical cell microscopy},
	url = {https://www.sciencedirect.com/science/article/pii/S2451910320300995},
	doi = {10.1016/j.coelec.2020.04.018},
	urldate = {2026-07-03},
	journal = {Current Opinion in Electrochemistry},
	author = {Wahab, Oluwasegun J. and Kang, Minkyung and Unwin, Patrick R.},
	month = aug,
	year = {2020},
	pages = {120--128},
}

@article{aaronson_scanning_2015,
	title = {Scanning {Electrochemical} {Cell} {Microscopy} {Platform} for {Ultrasensitive} {Photoelectrochemical} {Imaging}},
	volume = {87},
	issn = {0003-2700},
	url = {https://doi.org/10.1021/acs.analchem.5b00288},
	doi = {10.1021/acs.analchem.5b00288},
	number = {8},
	urldate = {2026-07-03},
	journal = {Analytical Chemistry},
	publisher = {American Chemical Society},
	author = {Aaronson, Barak D. B. and Byers, Joshua C. and Colburn, Alex W. and McKelvey, Kim and Unwin, Patrick R.},
	month = apr,
	year = {2015},
	pages = {4129--4133},
}

@article{mcbrayer_scanning_2024,
	title = {Scanning {Electrochemical} {Microscopy} {Reveals} {That} {Model} {Silicon} {Anodes} {Demonstrate} {Global} {Solid} {Electrolyte} {Interphase} {Passivation} {Degradation} during {Calendar} {Aging}},
	volume = {16},
	issn = {1944-8244},
	url = {https://doi.org/10.1021/acsami.3c14361},
	doi = {10.1021/acsami.3c14361},
	number = {15},
	urldate = {2026-07-02},
	journal = {ACS Applied Materials \& Interfaces},
	publisher = {American Chemical Society},
	author = {McBrayer, Josefine D. and Schorr, Noah B. and Lam, Mila Nhu and Meyerson, Melissa L. and Harrison, Katharine L. and Minteer, Shelley D.},
	month = apr,
	year = {2024},
	pages = {19663--19671},
}

@article{wang_theory_1993,
	title = {Theory and applications of guided-mode resonance filters},
	volume = {32},
	copyright = {© 1993 Optical Society of America},
	issn = {2155-3165},
	url = {https://opg.optica.org/ao/abstract.cfm?uri=ao-32-14-2606},
	doi = {10.1364/AO.32.002606},
	language = {EN},
	number = {14},
	urldate = {2026-07-02},
	journal = {Applied Optics},
	publisher = {Optica Publishing Group},
	author = {Wang, S. S. and Magnusson, R.},
	month = may,
	year = {1993},
	pages = {2606--2613},
}

@article{knief_disorder_1999,
	title = {Disorder, defects, and optical absorption in a − {Si} and a − {S} i : {H}},
	volume = {59},
	copyright = {http://link.aps.org/licenses/aps-default-license},
	issn = {0163-1829, 1095-3795},
	shorttitle = {Disorder, defects, and optical absorption in a − {Si} and a − {S} i},
	url = {https://link.aps.org/doi/10.1103/PhysRevB.59.12940},
	doi = {10.1103/PhysRevB.59.12940},
	language = {en},
	number = {20},
	urldate = {2026-07-01},
	journal = {Physical Review B},
	author = {Knief, Simone and Von Niessen, Wolfgang},
	month = may,
	year = {1999},
	pages = {12940--12946},
}

@article{esser_ultrafast_1990,
	title = {Ultrafast recombination and trapping in amorphous silicon},
	volume = {41},
	doi = {10.1103/PhysRevB.41.2879},
	number = {5},
	journal = {Physical Review B},
	author = {Esser, A.},
	year = {1990},
	pages = {2879--2884},
}

@article{choi_sn-coupled_2014,
	title = {Sn-{Coupled} p-{Si} {Nanowire} {Arrays} for {Solar} {Formate} {Production} from {CO2}},
	volume = {4},
	copyright = {© 2014 WILEY-VCH Verlag GmbH \& Co. KGaA, Weinheim},
	issn = {1614-6840},
	url = {https://onlinelibrary.wiley.com/doi/abs/10.1002/aenm.201301614},
	doi = {10.1002/aenm.201301614},
	number = {11},
	urldate = {2026-06-30},
	journal = {Advanced Energy Materials},
	author = {Choi, Sung Kyu and Kang, Unseock and Lee, Seunghoon and Ham, Dong Jin and Ji, Sang Min and Park, Hyunwoong},
	year = {2014},
	note = {\_eprint: https://advanced.onlinelibrary.wiley.com/doi/pdf/10.1002/aenm.201301614},
	pages = {1301614},
}

@article{boettcher_photoelectrochemical_2011,
	title = {Photoelectrochemical {Hydrogen} {Evolution} {Using} {Si} {Microwire} {Arrays}},
	volume = {133},
	issn = {0002-7863},
	url = {https://doi.org/10.1021/ja108801m},
	doi = {10.1021/ja108801m},
	number = {5},
	urldate = {2026-06-30},
	journal = {Journal of the American Chemical Society},
	publisher = {American Chemical Society},
	author = {Boettcher, Shannon W. and Warren, Emily L. and Putnam, Morgan C. and Santori, Elizabeth A. and Turner-Evans, Daniel and Kelzenberg, Michael D. and Walter, Michael G. and McKone, James R. and Brunschwig, Bruce S. and Atwater, Harry A. and Lewis, Nathan S.},
	month = feb,
	year = {2011},
	pages = {1216--1219},
}

@article{bell_light_2013,
	title = {Light intensity effects on photocatalytic water splitting with a titania catalyst},
	volume = {38},
	issn = {0360-3199},
	url = {https://www.sciencedirect.com/science/article/pii/S0360319913007714},
	doi = {10.1016/j.ijhydene.2013.02.147},
	number = {17},
	urldate = {2026-06-21},
	journal = {International Journal of Hydrogen Energy},
	author = {Bell, Stuart and Will, Geoffrey and Bell, John},
	month = jun,
	year = {2013},
	pages = {6938--6947},
}

@article{ghoshal_catalyst-free_2019,
	title = {Catalyst-{Free} and {Morphology}-{Controlled} {Growth} of {2D} {Perovskite} {Nanowires} for {Polarized} {Light} {Detection}},
	volume = {7},
	copyright = {© 2019 WILEY-VCH Verlag GmbH \& Co. KGaA, Weinheim},
	issn = {2195-1071},
	url = {https://onlinelibrary.wiley.com/doi/abs/10.1002/adom.201900039},
	doi = {10.1002/adom.201900039},
	language = {en},
	number = {15},
	urldate = {2026-06-21},
	journal = {Advanced Optical Materials},
	author = {Ghoshal, Debjit and Wang, Tianmeng and Tsai, Hsin-Zon and Chang, Shao-Wen and Crommie, Michael and Koratkar, Nikhil and Shi, Su-Fei},
	year = {2019},
	note = {\_eprint: https://advanced.onlinelibrary.wiley.com/doi/pdf/10.1002/adom.201900039},
	pages = {1900039},
}

@article{christopher_singular_2012,
	title = {Singular characteristics and unique chemical bond activation mechanisms of photocatalytic reactions on plasmonic nanostructures},
	volume = {11},
	copyright = {2012 Springer Nature Limited},
	issn = {1476-4660},
	url = {https://www.nature.com/articles/nmat3454},
	doi = {10.1038/nmat3454},
	language = {en},
	number = {12},
	urldate = {2026-06-21},
	journal = {Nature Materials},
	publisher = {Nature Publishing Group},
	author = {Christopher, Phillip and Xin, Hongliang and Marimuthu, Andiappan and Linic, Suljo},
	month = dec,
	year = {2012},
	pages = {1044--1050},
}

@article{chen_origin_2023,
	title = {Origin of {Superlinear} {Power} {Dependence} of {Reaction} {Rates} in {Plasmon}-{Driven} {Photocatalysis}: {A} {Case} {Study} of {Reductive} {Nitrothiophenol} {Coupling} {Reactions}},
	volume = {23},
	copyright = {https://doi.org/10.15223/policy-029},
	issn = {1530-6984, 1530-6992},
	shorttitle = {Origin of {Superlinear} {Power} {Dependence} of {Reaction} {Rates} in {Plasmon}-{Driven} {Photocatalysis}},
	url = {https://pubs.acs.org/doi/10.1021/acs.nanolett.3c00195},
	doi = {10.1021/acs.nanolett.3c00195},
	language = {en},
	number = {7},
	urldate = {2026-06-21},
	journal = {Nano Letters},
	author = {Chen, Kexun and Wang, Hui},
	month = apr,
	year = {2023},
	pages = {2870--2876},
}

@article{verma_paradox_2024,
	title = {The paradox of thermal vs. non-thermal effects in plasmonic photocatalysis},
	volume = {15},
	copyright = {2024 The Author(s)},
	issn = {2041-1723},
	url = {https://www.nature.com/articles/s41467-024-51916-3},
	doi = {10.1038/s41467-024-51916-3},
	language = {en},
	number = {1},
	urldate = {2026-06-21},
	journal = {Nature Communications},
	publisher = {Nature Publishing Group},
	author = {Verma, Rishi and Sharma, Gunjan and Polshettiwar, Vivek},
	month = sep,
	year = {2024},
	pages = {7974},
}

@article{huttenhofer_metasurface_2021,
	title = {Metasurface {Photoelectrodes} for {Enhanced} {Solar} {Fuel} {Generation}},
	volume = {11},
	copyright = {© 2021 The Authors. Advanced Energy Materials published by Wiley-VCH GmbH},
	issn = {1614-6840},
	url = {https://onlinelibrary.wiley.com/doi/abs/10.1002/aenm.202102877},
	doi = {10.1002/aenm.202102877},
	language = {en},
	number = {46},
	urldate = {2026-06-20},
	journal = {Advanced Energy Materials},
	author = {Hüttenhofer, Ludwig and Golibrzuch, Matthias and Bienek, Oliver and Wendisch, Fedja J. and Lin, Rui and Becherer, Markus and Sharp, Ian D. and Maier, Stefan A. and Cortés, Emiliano},
	year = {2021},
	note = {\_eprint: https://advanced.onlinelibrary.wiley.com/doi/pdf/10.1002/aenm.202102877},
	pages = {2102877},
}

@article{liu_bifacial_2020,
	title = {Bifacial passivation of n-silicon metal–insulator–semiconductor photoelectrodes for efficient oxygen and hydrogen evolution reactions},
	volume = {13},
	issn = {1754-5706},
	url = {https://pubs.rsc.org/en/content/articlelanding/2020/ee/c9ee02766a},
	doi = {10.1039/C9EE02766A},
	language = {en},
	number = {1},
	urldate = {2026-06-20},
	journal = {Energy \& Environmental Science},
	publisher = {The Royal Society of Chemistry},
	author = {Liu, Bin and Feng, Shijia and Yang, Lifei and Li, Chengcheng and Luo, Zhibin and Wang, Tuo and Gong, Jinlong},
	month = jan,
	year = {2020},
	pages = {221--228},
}

@article{peng_progress_2025,
	title = {Progress on {Si}-based photoelectrodes for industrial production of green hydrogen by solar-driven water splitting},
	volume = {3},
	copyright = {© 2024 The Author(s). EcoEnergy published by John Wiley \& Sons Australia, Ltd on behalf of China Chemical Safety Association.},
	issn = {2835-9399},
	url = {https://onlinelibrary.wiley.com/doi/abs/10.1002/ece2.73},
	doi = {10.1002/ece2.73},
	language = {en},
	number = {1},
	urldate = {2026-06-20},
	journal = {EcoEnergy},
	author = {Peng, Shuyang and Liu, Di and Bai, Haoyun and Liu, Chunfa and Feng, Jinxian and An, Keyu and Qiao, Lulu and Lo, Kin Ho and Pan, Hui},
	year = {2025},
	note = {\_eprint: https://onlinelibrary.wiley.com/doi/pdf/10.1002/ece2.73},
	pages = {25--55},
}

@article{roh_photoelectrochemical_2022,
	title = {Photoelectrochemical {CO2} {Reduction} toward {Multicarbon} {Products} with {Silicon} {Nanowire} {Photocathodes} {Interfaced} with {Copper} {Nanoparticles}},
	volume = {144},
	issn = {0002-7863},
	url = {https://doi.org/10.1021/jacs.2c03702},
	doi = {10.1021/jacs.2c03702},
	number = {18},
	urldate = {2026-06-18},
	journal = {Journal of the American Chemical Society},
	publisher = {American Chemical Society},
	author = {Roh, Inwhan and Yu, Sunmoon and Lin, Chung-Kuan and Louisia, Sheena and Cestellos-Blanco, Stefano and Yang, Peidong},
	month = may,
	year = {2022},
	pages = {8002--8006},
}

@misc{padhy_temperature_2025,
	title = {Temperature bandgaps and thermal dopants arising from photothermal nonlinearities in high-{Q} silicon metasurfaces},
	url = {http://arxiv.org/abs/2511.12038},
	doi = {10.48550/arXiv.2511.12038},
	urldate = {2026-06-18},
	publisher = {arXiv},
	author = {Padhy, Punnag and Zaman, Mohammad Asif and Dionne, Jennifer},
	month = nov,
	year = {2025},
	note = {arXiv:2511.12038 [physics.optics]},
}

@article{lee_scalable_2021,
	title = {Scalable, highly stable {Si}-based metal-insulator-semiconductor photoanodes for water oxidation fabricated using thin-film reactions and electrodeposition},
	volume = {12},
	copyright = {2021 The Author(s)},
	issn = {2041-1723},
	url = {https://www.nature.com/articles/s41467-021-24229-y},
	doi = {10.1038/s41467-021-24229-y},
	language = {en},
	number = {1},
	urldate = {2026-06-18},
	journal = {Nature Communications},
	publisher = {Nature Publishing Group},
	author = {Lee, Soonil and Ji, Li and De Palma, Alex C. and Yu, Edward T.},
	month = jun,
	year = {2021},
	pages = {3982},
}

@article{pahlevaninezhad_nano-optic_2018,
	title = {Nano-optic endoscope for high-resolution optical coherence tomography in vivo},
	volume = {12},
	copyright = {2018 The Author(s)},
	issn = {1749-4893},
	url = {https://www.nature.com/articles/s41566-018-0224-2},
	doi = {10.1038/s41566-018-0224-2},
	language = {en},
	number = {9},
	urldate = {2026-06-18},
	journal = {Nature Photonics},
	publisher = {Nature Publishing Group},
	author = {Pahlevaninezhad, Hamid and Khorasaninejad, Mohammadreza and Huang, Yao-Wei and Shi, Zhujun and Hariri, Lida P. and Adams, David C. and Ding, Vivien and Zhu, Alexander and Qiu, Cheng-Wei and Capasso, Federico and Suter, Melissa J.},
	month = sep,
	year = {2018},
	pages = {540--547},
}

@article{wang_resonantly_2024,
	title = {Resonantly enhanced second- and third-harmonic generation in dielectric nonlinear metasurfaces},
	volume = {7},
	copyright = {http://creativecommons.org/licenses/by/3.0/},
	issn = {2096-4579},
	url = {https://www.oejournal.org/oea/en/article/doi/10.29026/oea.2024.230186.pdf},
	doi = {10.29026/oea.2024.230186},
	language = {en},
	number = {5},
	urldate = {2026-06-18},
	journal = {Opto-Electronic Advances},
	author = {Wang, Ji Tong and Tonkaev, Pavel and Koshelev, Kirill and Lai, Fangxing and Kruk, Sergey and Song, Qinghai and Kivshar, Yuri and Panoiu, Nicolae C.},
	month = may,
	year = {2024},
	pages = {230186--15},
}

@article{song_laterally_2014,
	title = {Laterally assembled nanowires for ultrathin broadband solar absorbers},
	volume = {22},
	copyright = {© 2014 Optical Society of America},
	issn = {1094-4087},
	url = {https://opg.optica.org/oe/abstract.cfm?uri=oe-22-S3-A992},
	doi = {10.1364/OE.22.00A992},
	language = {EN},
	number = {103},
	urldate = {2026-06-18},
	journal = {Optics Express},
	publisher = {Optica Publishing Group},
	author = {Song, Kyung-Deok and Kempa, Thomas J. and Park, Hong-Gyu and Kim, Sun-Kyung},
	month = may,
	year = {2014},
	pages = {A992--A1000},
}

@article{hale_perfect_2020,
	title = {Perfect absorption in {GaAs} metasurfaces near the bandgap edge},
	volume = {28},
	issn = {1094-4087},
	url = {https://opg.optica.org/oe/abstract.cfm?uri=oe-28-23-35284},
	doi = {10.1364/OE.404249},
	language = {EN},
	number = {23},
	urldate = {2026-06-18},
	journal = {Optics Express},
	publisher = {Optica Publishing Group},
	author = {Hale, L. L. and Vabishchevich, P. P. and Siday, T. and Harris, C. T. and Luk, T. S. and Addamane, S. J. and Reno, J. L. and Brener, I. and Mitrofanov, O.},
	month = nov,
	year = {2020},
	pages = {35284--35296},
}

@article{lin_photochemical_2023,
	title = {Photochemical {Diodes} for {Simultaneous} {Bias}-{Free} {Glycerol} {Valorization} and {Hydrogen} {Evolution}},
	volume = {145},
	issn = {0002-7863},
	url = {https://doi.org/10.1021/jacs.3c01982},
	doi = {10.1021/jacs.3c01982},
	number = {24},
	urldate = {2026-06-17},
	journal = {Journal of the American Chemical Society},
	publisher = {American Chemical Society},
	author = {Lin, Jia-An and Roh, Inwhan and Yang, Peidong},
	month = jun,
	year = {2023},
	pages = {12987--12991},
}

@incollection{kilner_j_a_photovoltaic_2012,
	title = {Photovoltaic ({PV}) thin-films for solar cells},
	isbn = {978-0-85709-059-1},
	url = {https://www.sciencedirect.com/book/edited-volume/9780857090591/functional-materials-for-sustainable-energy-applications},
	urldate = {2026-05-03},
	booktitle = {Functional {Materials} for {Sustainable} {Energy} {Applications}},
	publisher = {Woodhead Publishing Series in Energy},
	author = {{Kilner, J. A.} and {Skinner, Stephen J.} and {Irvine, Stuart J.C.} and {Edwards, Peter P.}},
	year = {2012},
	note = {ISBN: 9780857090591},
	pages = {22--41},
}

@article{segev_2022_2022,
	title = {The 2022 solar fuels roadmap},
	issn = {0022-3727},
	url = {https://depositonce.tu-berlin.de/handle/11303/17188},
	language = {en},
	urldate = {2026-05-03},
	author = {Segev, Gideon and Kibsgaard, Jakob and Hahn, Christopher and Xu, Zhichuan J. and Cheng, Wen-Hui (Sophia) and Deutsch, Todd G. and Xiang, Chengxiang and Zhang, Jenny Z. and Hammarström, Leif and Nocera, Daniel G. and Weber, Adam Z. and Agbo, Peter and Hisatomi, Takashi and Osterloh, Frank E. and Domen, Kazunari and Abdi, Fatwa F. and Haussener, Sophia and Miller, Daniel J. and Ardo, Shane and McIntyre, Paul C. and Hannappel, Thomas and Hu, Shu and Atwater, Harry and Gregoire, John M. and Ertem, Mehmed Z. and Sharp, Ian D. and Choi, Kyoung-Shin and Lee, Jae Sung and Ishitani, Osamu and Ager, Joel W. and Prabhakar, Rajiv Ramanujam and Bell, Alexis T. and Boettcher, Shannon W. and Vincent, Kylie and Takanabe, Kazuhiro and Artero, Vincent and Napier, Ryan and Roldán Cuenya, Beatriz and Koper, Marc T. M. and Van De Krol, Roel and Houle, Frances},
	month = jun,
	year = {2022},
}

@article{andrei_nanowire_2023,
	title = {Nanowire photochemical diodes for artificial photosynthesis},
	volume = {9},
	url = {https://www.science.org/doi/10.1126/sciadv.ade9044},
	doi = {10.1126/sciadv.ade9044},
	number = {6},
	urldate = {2026-05-03},
	journal = {Science Advances},
	publisher = {American Association for the Advancement of Science},
	author = {Andrei, Virgil and Roh, Inwhan and Yang, Peidong},
	month = feb,
	year = {2023},
	pages = {eade9044},
}

@article{chen_atomic_2011,
	title = {Atomic layer-deposited tunnel oxide stabilizes silicon photoanodes for water oxidation},
	volume = {10},
	copyright = {2011 Springer Nature Limited},
	issn = {1476-4660},
	url = {https://www.nature.com/articles/nmat3047},
	doi = {10.1038/nmat3047},
	language = {en},
	number = {7},
	urldate = {2026-05-03},
	journal = {Nature Materials},
	publisher = {Nature Publishing Group},
	author = {Chen, Yi Wei and Prange, Jonathan D. and Dühnen, Simon and Park, Yohan and Gunji, Marika and Chidsey, Christopher E. D. and McIntyre, Paul C.},
	month = jul,
	year = {2011},
	pages = {539--544},
}

@article{karaman_photo-thermally_2026,
	title = {Photo-{Thermally} {Tunable} {Photon}-{Pair} {Generation} in {Dielectric} {Metasurfaces}},
	volume = {20},
	issn = {1936-0851},
	url = {https://doi.org/10.1021/acsnano.5c14740},
	doi = {10.1021/acsnano.5c14740},
	number = {5},
	urldate = {2026-04-30},
	journal = {ACS Nano},
	publisher = {American Chemical Society},
	author = {Karaman, Omer Can and Li, Hua and Dayi, Elif Nur and Galland, Christophe and Tagliabue, Giulia},
	month = feb,
	year = {2026},
	pages = {4079--4087},
}

@article{liu_geometry-programmable_2026,
	title = {Geometry-{Programmable} {Light}-{Driven} {Silicon} {Microrobots}},
	volume = {38},
	issn = {1521-4095},
	url = {https://onlinelibrary.wiley.com/doi/abs/10.1002/adma.202522384},
	doi = {10.1002/adma.202522384},
	language = {en},
	number = {20},
	urldate = {2026-04-30},
	journal = {Advanced Materials},
	author = {Liu, Kunfeng and Huang, Rong and Li, Wanyuan and Yao, Jingsong and Lei, Dapeng and Yang, Guangdong and Huang, Zhuochen and Chen, Li and Sun, Hao and Xiong, Ze and Wang, Jizhuang and Tang, Jinyao and Li, Dan},
	year = {2026},
	note = {\_eprint: https://advanced.onlinelibrary.wiley.com/doi/pdf/10.1002/adma.202522384},
	pages = {e22384},
}

@article{kiani_distinguishing_2024,
	title = {Distinguishing {Inner} and {Outer}-{Sphere} {Hot} {Electron} {Transfer} in {Au}/p-{GaN} {Photocathodes}},
	volume = {24},
	issn = {1530-6984},
	url = {https://doi.org/10.1021/acs.nanolett.4c04319},
	doi = {10.1021/acs.nanolett.4c04319},
	number = {50},
	urldate = {2026-03-20},
	journal = {Nano Letters},
	publisher = {American Chemical Society},
	author = {Kiani, Fatemeh and Bowman, Alan R. and Sabzehparvar, Milad and Sundararaman, Ravishankar and Tagliabue, Giulia},
	month = dec,
	year = {2024},
	pages = {16008--16014},
}

@article{kiani_transport_2023,
	title = {Transport and {Interfacial} {Injection} of d-{Band} {Hot} {Holes} {Control} {Plasmonic} {Chemistry}},
	volume = {8},
	url = {https://doi.org/10.1021/acsenergylett.3c01505},
	doi = {10.1021/acsenergylett.3c01505},
	number = {10},
	urldate = {2025-08-14},
	journal = {ACS Energy Letters},
	publisher = {American Chemical Society},
	author = {Kiani, Fatemeh and Bowman, Alan R. and Sabzehparvar, Milad and Karaman, Can O. and Sundararaman, Ravishankar and Tagliabue, Giulia},
	month = oct,
	year = {2023},
	pages = {4242--4250},
}

@article{cortes_optical_2022,
	title = {Optical {Metasurfaces} for {Energy} {Conversion}},
	volume = {122},
	issn = {0009-2665},
	url = {https://doi.org/10.1021/acs.chemrev.2c00078},
	doi = {10.1021/acs.chemrev.2c00078},
	number = {19},
	urldate = {2025-07-16},
	journal = {Chemical Reviews},
	publisher = {American Chemical Society},
	author = {Cortés, Emiliano and Wendisch, Fedja J. and Sortino, Luca and Mancini, Andrea and Ezendam, Simone and Saris, Seryio and de S. Menezes, Leonardo and Tittl, Andreas and Ren, Haoran and Maier, Stefan A.},
	month = oct,
	year = {2022},
	pages = {15082--15176},
}

@article{paudel_metasurface-enhanced_2024,
	title = {Metasurface-enhanced photochemical activity in visible light absorbing semiconductors},
	volume = {160},
	issn = {0021-9606},
	url = {https://doi.org/10.1063/5.0199589},
	doi = {10.1063/5.0199589},
	number = {14},
	urldate = {2024-11-25},
	journal = {The Journal of Chemical Physics},
	author = {Paudel, Yamuna and Chachayma-Farfan, Diego J. and Alù, Andrea and Sfeir, Matthew Y.},
	month = apr,
	year = {2024},
	pages = {144710},
}
